\documentclass[a4paper,11pt]{article}
\pdfoutput=0 
\usepackage{jheppub} 
\usepackage{subfig}
\usepackage[T1]{fontenc} 
\usepackage{graphicx}
\usepackage{epsfig}
\usepackage{axodraw}
\usepackage{amsmath}
\usepackage{amssymb}
\usepackage{float}
\usepackage{placeins}

\newcommand{\be}{\begin{equation}}
\newcommand{\ee}{\end{equation}}
\newcommand{\nn}{\nonumber}
\newcommand{\bea}{\begin{eqnarray}}
\newcommand{\eea}{\end{eqnarray}}
\newcommand{\bfig}{\begin{figure}}
\newcommand{\efig}{\end{figure}}
\newcommand{\bc}{\begin{center}}
\newcommand{\ec}{\end{center}}
\newcommand{\ep}{\epsilon}

\newcommand{\Li}{\text{Li}}

\allowdisplaybreaks[4]

\title{\boldmath Two-loop planar master integrals for Higgs$\to 3$ partons with full heavy-quark mass dependence}

\preprint{MITP/16-088}

\author[a,b]{Roberto Bonciani,}\author[c,d]{Vittorio Del Duca,}\author[e]{Hjalte Frellesvig,}\author[f]{Johannes M. Henn,}\author[a,b,c]{Francesco Moriello,}\author[g]{Vladimir A. Smirnov}

\affiliation[a]{Sapienza - Universit\`a di Roma, Dipartimento di Fisica, Piazzale Aldo Moro 5, 00185, Rome, Italy}

\affiliation[b]{INFN Sezione di Roma, Piazzale Aldo Moro 2, 00185, Rome, Italy}

\affiliation[c]{ETH Zurich, Institut fur theoretische Physik, Wolfgang-Paulistr. 27, 8093, Zurich, Switzerland}

\affiliation[d]{INFN Laboratori Nazionali di Frascati, 00044 Frascati (Roma), Italy}

\affiliation[e]{Institute of Nuclear and Particle Physics, NCSR Demokritos,
Agia Paraskevi, 15310, Greece}

\affiliation[f]{PRISMA Cluster of Excellence, Johannes Gutenberg University, 55099 Mainz, Germany}

\affiliation[g]{Skobeltsyn Inst. of Nuclear Physics of Moscow State University, 119991 Moscow, Russia}

\emailAdd{roberto.bonciani@roma1.infn.it}
\emailAdd{delducav@itp.phys.ethz.ch}
\emailAdd{frellesvig@inp.demokritos.gr}
\emailAdd{henn@uni-mainz.de}
\emailAdd{fmoriell@phys.ethz.ch}
\emailAdd{smirnov@theory.sinp.msu.ru}

\abstract{We present the analytic computation of all the planar master integrals which contribute to the two-loop scattering amplitudes for Higgs$\to 3$ partons, with full heavy-quark mass dependence. These are relevant for the NNLO corrections to fully inclusive Higgs production and to the NLO corrections to Higgs production in association with a jet, in the full theory. The computation is performed 
using the differential equations method. Whenever possible, a basis of master integrals that are pure functions of uniform weight is used. The result is expressed in terms of one-fold integrals of polylogarithms and elementary functions up to transcendental weight four. Two integral sectors are expressed in terms of elliptic functions. We show that by introducing a one-dimensional parametrization of the integrals the relevant second order differential equation can be readily solved, and the solution can be expressed to all orders of the dimensional regularization parameter in terms of iterated integrals over elliptic kernels.
We express the result for the elliptic sectors in terms of two and three-fold iterated integrals, which we find suitable for numerical evaluations. This is the first time that four-point multiscale Feynman integrals have been computed in a fully analytic way in terms of elliptic functions.
}

\begin{document} 
\maketitle
\flushbottom

\section{Introduction}
\label{sec:intro}

At the Large Hadron Collider (LHC), the main production mode of the Standard Model (SM) Higgs
boson is via gluon-gluon fusion. The Higgs boson does not couple directly to the gluons, the
interaction being mediated by a heavy-quark loop. That makes the evaluation of the radiative
corrections to Higgs boson production via gluon-gluon fusion challenging, since the Born
process is computed through one-loop diagrams, the next-to-leading order (NLO) QCD corrections
involve the computation of two-loop diagrams, the next-to-next-to-leading order (NNLO)
corrections the computation of three-loop diagrams, and so on. In fact, fully inclusive Higgs
production is known up to NLO~\cite{Graudenz:1992pv,Spira:1995rr}, while Higgs production in
association with one jet~\cite{Ellis:1987xu} and the Higgs $p_T$
distribution~\cite{Kauffman:1991jt} are known only at leading order.

The evaluation of the radiative corrections simplifies considerably in the Higgs effective
field theory (HEFT), where the heavy quark is integrated out and the Higgs boson couples
directly to the gluons, effectively reducing the computation by one loop. For fully inclusive
Higgs production, the HEFT is valid when the Higgs mass is smaller than the heavy-quark mass,
$m_H \lesssim m_Q$. Thus it is expected to be a good approximation to the full theory (FT),
which gets corrections from the top-mass contribution and from the top-bottom interference. In
fact, using the FT NLO computation as a benchmark, one can see that the HEFT NLO computation
approximates very well the FT NLO computation, since the top-bottom interference and the
top-mass corrections are about the same size although with opposite
sign~\cite{Grazzini:2013mca}. At NNLO, the FT mass corrections are expected to be in the
percent range, which is though competitive with the precision of the HEFT computation at
next-to-next-to-next-to-leading order (${\rm
N^3LO}$)~\cite{Anastasiou:2015ema,Anastasiou:2016cez}.

For Higgs production in association with one jet or for the Higgs $p_T$ distribution, using
the leading-order results~\cite{Ellis:1987xu,Kauffman:1991jt} as a benchmark one can show that
the HEFT is valid when $m_H \lesssim m_Q$ and the jet or Higgs transverse momenta are smaller
than the heavy-quark mass, $p_T \lesssim m_Q$~\cite{Baur:1989cm,deFlorian:1999zd}. In the
HEFT, Higgs production in association with one jet~\cite{Boughezal:2015dra,Boughezal:2015aha}
and the Higgs $p_T$ distribution~\cite{Chen:2016zka} are known at NNLO. No complete FT results
are known beyond the leading order. Approximate NLO top-mass effects have been computed, and
shown to be small and to agree well with the HEFT for  $p_T \lesssim
m_{top}$~\cite{Harlander:2012hf,Frederix:2016cnl,Caola:2016upw} and up to $p_T\sim
300$~GeV~\cite{Neumann:2016dny}. However, they are expected to be non-negligible in the high
$p_T$ tail. Finally, it is worth noting that in many New Physics (NP) models, the high $p_T$
tail of the Higgs $p_T$ distribution is sensitive to modifications of the Higgs-top
coupling~\cite{Grojean:2013nya,Azatov:2013xha,Azatov:2016xik}.


In this paper, we report on the analytic computation of all the planar master integrals which
are needed to compute the two-loop scattering amplitudes for Higgs$\to 3$ partons, with full
heavy-quark mass dependence. These are relevant to compute the FT NNLO corrections to fully
inclusive Higgs production and the FT NLO corrections to Higgs production in association with
one jet or to the Higgs $p_T$ distribution.

The differential equations
method~\cite{Kotikov:1990kg,Kotikov:1991pm,Bern:1993kr,Remiddi:1997ny,Gehrmann:1999as} has
proven to be one of the most powerful tools to compute (dimensionally regularized) loop
Feynman integrals. In particular, the reduction of the Feynman integrals to a set of linearly
independent integrals, dubbed master 
integrals~\cite{Tkachov:1981wb,Chetyrkin:1981qh,Laporta:1996mq,Laporta:2001dd}, through
integration-by-parts identities,  the exploration of new classes of special functions such as
multiple polylogarithms~\cite{Goncharov:1998kja,Remiddi:1999ew},  and a better understanding
of their functional properties~\cite{Goncharov:2010jf,Duhr:2011zq,Frellesvig:2016ske}, have
made the technique increasingly efficient. However, until recently the method was mostly
applied in relatively simple kinematic situations, with the Feynman integrals depending on few
scales, while complicated integrals needed a case-by-case analysis.

A major breakthrough was made in \cite{Henn:2013pwa}, where a canonical form of the
differential equations for Feynman integrals was proposed. A key idea is that the canonical
basis can be found by inspecting the singularity structure of the loop integrand. More
precisely, one computes the leading singularities, i.e. maximal multidimensional residues of
the loop integrand \cite{Cachazo:2008vp,ArkaniHamed:2010gh}. The fact that this can be done
before the differential equations are set up renders this technique extremely
efficient\footnote{An alternative approach to finding a canonical basis was proposed in
\cite{Henn:2014qga1,Henn:2014qga,Lee:2014ioa}. It is based on the idea of transforming the
system of differential equations such that the order of all singularities is manifest.  In
their current form, the ensuing algorithms require that the integrals depend in a rational way
on a single variable, and usually yield rather complicated transformation matrices.}. When
considering differential equations for a set of integrals defined to be pure functions of
uniform weight, all relevant information about the analytic properties of the result is
manifest at the level of the equations. Moreover, it is possible to find an analytic
expression for the master integrals in terms of iterated integrals over algebraic kernels in a
fully algorithmic way, up to any order of the dimensional regularization parameter
(see~\cite{Henn:2013woa,Henn:2013nsa,Argeri:2014qva,Henn:2014lfa,Gehrmann:2014bfa,Caola:2014lpa,
DiVita:2014pza,Bell:2014zya,Lee:2014ioa,Bonciani:2015eua,Gehrmann:2015dua,Grozin:2015kna,Gehrmann:2015bfy,Henn:2016men,Bonciani:2016ypc,Henn:2016jdu,Lee:2016lvq}
for many applications of these ideas).  It is important to note that these ideas also
streamline calculations whose output cannot be immediately written in terms of multiple
polylogarithms, but where Chen iterated integrals \cite{Chen-iterated} are the appropriate
special functions, see e.g. \cite{Caron-Huot:2014lda,Gehrmann:2015bfy}. This class of
functions will also be important in this paper.

Beyond Chen iterated integrals, there are cases where elliptic functions appear. This is
typically related to several equations being coupled in four dimensions, see e.g.
\cite{Tancredi:2015pta, Caron-Huot:2014lda}. The appearance of elliptic functions can be also
anticipated by inspecting the maximal cuts of the corresponding loop integrands
\cite{CaronHuot:2012ab}. In this case the precise form of the canonical basis is not yet
known, and presumably finding it will involve a generalization of the concept of leading
singularities.

Over the last two decades a lot of effort has been made to understand the analytic properties
of Feynman integrals which go beyond the multiple polylogarithms case, mostly related to the
so-called sunrise
diagram~\cite{Henn:2014qga,Caffo:1998du,Laporta:2004rb,Bloch:2013tra,Bloch:2014qca,Adams:2014vja,Adams:2015gva,
Adams:2015ydq,Bloch:2016izu,Remiddi:2016gno,Adams:2016xah}. However, to the best of our
knowledge, such a generalized class of Feynman integrals has not been used so far in a fully
analytic computation of a four-point multiscale scattering amplitude. In this paper, we
compute in the Euclidean region all the planar master integrals relevant for Higgs$\to 3$
partons, retaining the full heavy-quark mass dependence, which include two elliptic integral
sectors.

We write down the differential equations following the approach of \cite{Henn:2013pwa}. We
find that most integrals can be expressed in terms of Chen iterated integrals
\cite{Chen-iterated}. The corresponding function alphabet depends on three dimensionless
variables and contains $49$ letters, underlining the complexity of the problem. Having a fast
and reliable numerical evaluation in mind, we derive a representation of all functions up to
weight two in terms of logarithms and dilogarithms. Following \cite{Caron-Huot:2014lda}, this
allows us to write the weight three and four functions in terms of one-fold integral
representations. We find the latter suitable for numerical evaluation. We show that the two
remaining integral sectors involve elliptic functions.  We analyze the corresponding system of
coupled equations, and solve them in a suitable variable. An important tool is to reduce the
problem to a one-variable problem  (a similar strategy has been used
in~\cite{Papadopoulos:2014lla} to effectively rationalize the alphabet of multiscale
processes).  The solution at any order in $\epsilon$ can be expressed in terms of iterated
integrals involving elliptic kernels. We then show that using auxiliary bases and basis
shifts, the result for the elliptic sectors can be expressed in terms of two and three-fold
iterated integrals, which we find suitable for numerical evaluation.  

The outline of the paper is as follows. In section~\ref{sec:integral families} we briefly
discuss the reduction to the master integrals and the kinematics of the processes under
consideration. In section~\ref{sec:DE} we review the differential equations method in the
context of pure functions of uniform weight, i.e. the canonical basis approach.  In
section~\ref{sec:solution canonical} we show that when a canonical basis exists the solution
can be expressed to all orders of the dimensional regularization parameter in terms of
multiple polylogarithms, also when a rational parametrization of the alphabet is not
possible.  We derive a one-fold integral representation of the result up to weight four which
is suitable for fast and reliable numerical evaluation.  In section~\ref{sec:elliptic sectors}
we discuss in detail how to analytically solve the elliptic sectors in terms of iterated
integrals over elliptic kernels. In section~\ref{sec:symbols} we discuss the class of
functions used to represent the elliptic sectors.  In section \ref{sec:conclusions} we
conclude and discuss future directions. We also provide six appendices in which we collect
more details about the calculation. In appendix \ref{sec:appendix:integral basis} we write the
explicit expressions for the canonical form of the master integrals, or conversely for the
basis choice in the elliptic case. In appendix \ref{sec:appendix:precanonical} we show the 125
master integrals in the pre-canonical form. In appendix \ref{sec:appendix:alphabet}, we give
the alphabet for the master integrals. In appendix \ref{sec:appendix:li2} we list the
dilogarithms we used to express the master integrals at weight two. In
appendix  \ref{sec:appendix:integral representation} we give more details about the one-fold
integral representation in terms of which we express the master integrals not depending on
elliptic functions. Finally in appendix  \ref{sec:appendix:cut} we show that the maximal cut
of the six-denominator elliptic sector provides useful information about the class of
functions  which characterise the sector.

\section{Notations and conventions}
\label{sec:integral families}

The leading order QCD contribution to Higgs decay to three partons, or alternatively to Higgs
production in hadronic collisions, is a process mediated by a loop of heavy quarks. This is
due to the fact that the SM Higgs boson does not couple directly to massless particles. The
decay channels are $H \rightarrow g g g$ and  $H \rightarrow g q \bar{q}$; the production
channels are $g g \rightarrow gH $, $g q \rightarrow q H$ and $q \bar{q} \rightarrow gH $. 
%
\begin{figure}
\bc
\[ \vcenter{
\hbox{
  \begin{picture}(0,0)(0,0)
\SetScale{0.65}
  \SetWidth{1.0}
\Line(-50,30)(-30,30)
\Line(-50,-30)(-30,-30)
\Line(30,30)(50,30)
  \SetWidth{3.5}
\DashLine(30,-30)(50,-30){3}
\Line(-30,30)(30,30)
\Line(-30,-30)(30,-30)
\Line(-30,-30)(-30,30)
\Line(30,-30)(30,30)
%
%
%
\end{picture}}
}
\]
\vspace*{5mm}
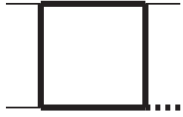
\caption{\label{figLO} Four-denominator topology for the LO contribution to the cross section
of Higgs boson production in association with a jet. Thick lines represent heavy quarks
propagators. Thin lines represent  massless external particles and propagators. The dashed
external line represents the Higgs boson.}
\ec
\efig
%
The one-loop Feynman diagrams for all these processes can be described using the
four-denominator {\it topology}\footnote{A topology is composed of the integrals for which the
same set of propagators have positive powers, while a \emph{subtopology} is a set of integrals
for which the propagators with positive powers are a subset of the ones of a given topology.}
(and {\it subtopologies}) depicted in fig.~\ref{figLO}.

\begin{figure}
\bc
\[ \vcenter{
\hbox{
  \begin{picture}(0,0)(0,0)
\SetScale{0.65}
  \SetWidth{1.0}
\Line(-50,30)(-30,30)
\Line(30,30)(50,30)
\Line(-50,-30)(30,-30)
\Line(-30,-30)(-30,30)
\Line(30,-30)(30,30)
\Line(0,-30)(0,30)
  \SetWidth{3.5}
\DashLine(30,-30)(50,-30){3}
\Line(-30,30)(30,30)
\Line(-30,-30)(30,-30)
\Line(-30,-30)(-30,30)
\Line(30,-30)(30,30)
%
%
\end{picture}}
}
\hspace{3.5cm}
\vcenter{
\hbox{
  \begin{picture}(0,0)(0,0)
\SetScale{0.65}
  \SetWidth{1.0}
\Line(-50,30)(0,30)
\Line(30,30)(50,30)
\Line(-50,-30)(0,-30)
\Line(-30,-30)(-30,30)
\Line(30,-30)(30,30)
\Line(0,-30)(0,30)
  \SetWidth{3.5}
\DashLine(30,-30)(50,-30){3}
\Line(0,30)(30,30)
\Line(0,-30)(30,-30)
\Line(30,-30)(30,30)
\Line(0,-30)(0,30)
%
%
\end{picture}}
}
\hspace{3.5cm}
\vcenter{
\hbox{
  \begin{picture}(0,0)(0,0)
\SetScale{0.65}
  \SetWidth{1.0}
\Line(-50,-30)(0,-30)
\Line(-50,30)(50,30)
\Line(-30,30)(-30,-30)
\Line(30,30)(30,-30)
  \SetWidth{3.5}
\DashLine(30,-30)(50,-30){3}
\Line(30,0)(0,-30)
\Line(0,-30)(30,-30)
\Line(30,0)(30,-30)
%
%
\end{picture}}
}
\hspace{3.5cm}
\vcenter{
\hbox{
  \begin{picture}(0,0)(0,0)
\SetScale{0.65}
  \SetWidth{1.0}
\Line(-50,-30)(0,-30)
\Line(-50,30)(50,30)
\Line(30,0)(0,-30)
  \SetWidth{3.5}
\DashLine(30,-30)(50,-30){3}
\Line(-30,-30)(30,-30)
\Line(-30,30)(30,30)
\Line(30,0)(30,-30)
\Line(-30,30)(-30,-30)
\Line(30,30)(30,-30)
%
%
\end{picture}}
}
\]
\vspace*{1.0cm}
\[ 
\vcenter{
\hbox{
  \begin{picture}(0,0)(0,0)
\SetScale{0.65}
  \SetWidth{1.0}
\Line(-50,-30)(0,-30)
\Line(-50,30)(50,30)
\Line(30,0)(0,30)
  \SetWidth{3.5}
\DashLine(30,-30)(50,-30){3}
\Line(-30,-30)(30,-30)
\Line(-30,30)(30,30)
\Line(30,0)(30,-30)
\Line(-30,30)(-30,-30)
\Line(30,30)(30,-30)
%
%
\end{picture}}
}
\hspace{3.5cm}
\vcenter{
\hbox{
  \begin{picture}(0,0)(0,0)
\SetScale{0.65}
  \SetWidth{1.0}
\Line(-50,-30)(0,-30)
\Line(-50,30)(50,30)
\Line(-30,0)(0,30)
  \SetWidth{3.5}
\DashLine(30,-30)(50,-30){3}
\Line(-30,-30)(30,-30)
\Line(-30,30)(30,30)
\Line(30,0)(30,-30)
\Line(-30,30)(-30,-30)
\Line(30,30)(30,-30)
%
%
\end{picture}}
}
\hspace{3.5cm}
\vcenter{
\hbox{
  \begin{picture}(0,0)(0,0)
\SetScale{0.65}
  \SetWidth{1.0}
\Line(-50,-30)(0,-30)
\Line(-50,30)(50,30)
\Line(0,30)(30,30)
\Line(30,30)(30,0)
  \SetWidth{3.5}
\DashLine(30,-30)(50,-30){3}
\Line(30,0)(0,30)
\Line(-30,-30)(30,-30)
\Line(-30,30)(0,30)
\Line(30,0)(30,-30)
\Line(-30,30)(-30,-30)
\Line(30,0)(30,-30)
%
%
\end{picture}}
}
\hspace{3.5cm}
\vcenter{
\hbox{
  \begin{picture}(0,0)(0,0)
\SetScale{0.65}
  \SetWidth{1.0}
\Line(-50,-30)(0,-30)
\Line(-50,30)(50,30)
\Line(-30,30)(0,30)
\Line(-30,30)(-30,0)
  \SetWidth{3.5}
\DashLine(30,-30)(50,-30){3}
\Line(-30,0)(0,30)
\Line(-30,-30)(30,-30)
\Line(0,30)(30,30)
\Line(30,0)(30,-30)
\Line(-30,0)(-30,-30)
\Line(30,30)(30,-30)
%
%
\end{picture}}
}
\]
\vspace*{5mm}
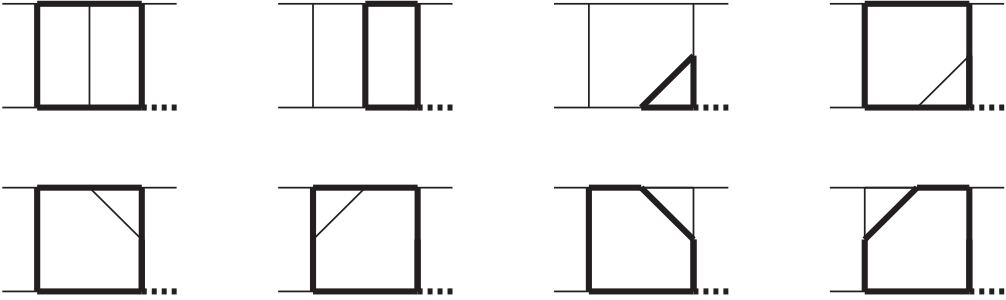
\caption{\label{figNLO} Planar seven-denominator topologies for the NLO contribution to the cross section of Higgs boson production in association with a jet in proton collisions, with full heavy-quark mass dependence.}
\ec
\efig
%

At NLO in $\alpha_S$, Feynman diagrams with up to seven propagators contribute to the processes above. They can all be described using the eight different planar seven-propagator topologies (and their subtopologies) depicted in fig.~\ref{figNLO}. 
We parametrized all eight topologies into nine-propagator integral families and we reduced 
the corresponding dimensionally regularized integrals to a minimal set of independent integrals, dubbed \emph{master integrals}, using the computer program FIRE~\cite{Smirnov:2008iw,Smirnov:2013dia,Smirnov:2014hma} combined with LiteRed~\cite{Lee:2012cn}. The list of denominators defining the integral families and additional details about this part of the calculation are provided in appendix \ref{sec:appendix:integral basis}.
%
\bfig
\bc
\[
\hspace*{-2mm}
 \vcenter{
\hbox{
}
}
\]
\vspace*{7mm}
\caption{Master integrals in pre-canonical form. Internal plain thin lines represent massless
propagators, while thick lines represent the top propagator. External plain thin lines
represent massless particles on their mass-shell. External dashed thin lines represent the
dependence on $s$, $t$, or $m_H^2$. The external dashed thick line represents the Higgs on its
mass-shell. The squared momentum $p^2$ can assume the values $p^2=s,t,m_H^2$. The squared
momentum $q^2$ can assume the values $q^2=s,m_H^2$. The squared momentum $r^2$ can assume the
values $r^2=s,t$. \label{figmasterprecan}}
\ec
\efig
%

The most general integral is defined in $D=4-2\epsilon$ space-time dimensions as,
\begin{equation}
I_{a_1,a_2,a_3,a_4,a_5,a_6,a_7,a_8,a_9}^{i}=\int \frac{d^D k_1 d^D k_2}{i \pi ^{D/2} i \pi ^{D/2}}\frac{[d_8^{i}]^{ -a_8}[d_9^{i}]^{- a_9}}{[d_1^{i}]^{ a_1}[d_2^{i}]^{ a_2}[d_3^{i}]^{ a_3}[d_4^{i}]^{ a_4}[d_5^{i}]^{ a_5}[d_6^{i}]^{ a_6}[d_7^{i}]^{ a_7}} \,,
\label{Integrals}
\end{equation}
where $i$ is the family index, and $a_i$ are integers. 
The reduction process leads to a set of 125 master integrals, shown in fig.~\ref{figmasterprecan}, that may be of relevance to more than
one physical process. We shall focus here on a Higgs boson decaying to three partons and on
Higgs$+$jet production. These processes differ by the physical phase-space region. Defining,
\be 
s=(p_1+p_2)^2, \quad\quad t=(p_1+p_3)^2, \quad\quad u=(p_2+p_3)^2, \quad\quad p_4^2 = s+t+u,
\ee  
where $p_1^2 = p_2^2 = p_3^2 = 0$, the relevant physical regions are
\be 
H\, {\rm decay} : \; s>0,\, t>0,\, u>0, \qquad H+{\rm jet} : \; s>p_4^2>0,\, t<0,\, u<0 \,,
\ee
both with the internal quark mass $m^2 > 0$. The integrals are functions of three dimensionless invariants,
\begin{equation}
x= \left\{ x_1,\,x_2,\,x_3 \right\},
\end{equation}
with
\begin{equation}
x_1= \frac{s}{m^2},\quad x_2=\frac{p_4^2}{m^2},\quad x_3=\frac{t}{m^2}.
\end{equation}

\FloatBarrier

In this paper we evaluate the integrals in the Euclidean region where no branch cuts are present, or rather in the subset there-of which has,
\begin{equation}
\label{eq:euclidean region}
x_3 < x_2 < x_1 < 0.
\end{equation}

It is then possible to analytically continue the result to the physical region using the Feynman prescription, by assigning a positive infinitesimal imaginary part to the external invariants and a negative infinitesimal imaginary part to the internal masses. The analytic continuation of the master integrals will be provided elsewhere.

The full basis of master integrals we evaluated in this paper is listed in appendix~\ref{sec:appendix:integral basis}.

The explicit results for the master integrals require about 200 MB to be stored in electronic form, and can be obtained upon request to the authors.

\section{Differential equations}
\label{sec:DE}

In order to analytically compute the master integrals we rely on the differential equations
method~\cite{Kotikov:1990kg,Bern:1993kr,Kotikov:1991pm,Remiddi:1997ny, Gehrmann:1999as}. All
the integrals discussed in this paper can be expressed in terms of multiple polylogarithms
except eight of them,
which involve elliptic functions.  In the polylogarithmic case we find a modified basis of
integrals that are pure functions of uniform weight~\cite{Henn:2013pwa}. In this basis the
differential equations take a canonical form and can be readily solved. This basis is found by
choosing integrals with constant leading singularities.  In the case of elliptic functions the
appropriate generalization of the notion of leading singularity has not yet been worked out.
It is nevertheless possible to choose a basis where the elliptic nature of the integrals is
manifest and the problem can be reduced to the solution of second order differential
equations, as we discuss in section \ref{sec:elliptic sectors}.

\subsection{General features of differential equations for Feynman integrals}

Denoting a set of $N$ basis integrals by $ f$, the set of kinematical variables by  $x$, and working in $D=4-2\epsilon$ dimensions, it is possible to define a system of first order linear differential equations for the integrals, that can be written in total generality as, 
\begin{equation}
\partial_m  f(x,\epsilon)= A_m(x, \epsilon) f(x,\epsilon)\,,
\end{equation}
where we used the shorthand $\partial_m =\partial/ \partial x_m$, and 
$A_m(x,\epsilon)$ is an $N\times N$ matrix with rational entries of its variables. The matrix $A_m(x,\epsilon)$ satisfies the integrability condition,
\begin{equation}
\partial_n A_m-\partial_m A_n - [A_n,A_m]=0\,,
\end{equation}
where $[A_n,A_m]= A_n A_m-A_m A_n$\,. 

The choice of the basis is not unique. Performing a basis change $f\rightarrow B  f$ the system of differential equations transforms according to 
\begin{equation}
A_m\rightarrow B^{-1}\partial_{x_m} B -B^{-1} A_m B\,.
\end{equation}
In~\cite{Henn:2013pwa} it was conjectured that performing a basis change with algebraic coefficients, for integral sectors expressible in terms of multiple polylogarithms, it is possible to factorize out the $\epsilon$ dependence of the differential equations,
\begin{equation}
\label{eq:def atilde}
\partial_m f(x,\epsilon)= \epsilon A_m(x) f(x, \epsilon)\,.
\end{equation}
Such a system of differential equations is said to be in canonical form. In order to discuss the properties of the solution it is convenient to write the differential equations in differential form,
\be
d f(x,\epsilon)= \epsilon \;d \tilde{A}(x) f(x, \epsilon),
\ee
where $\tilde{A}$ is a matrix such that,
\be 
\frac{\partial \tilde{A}(x)}{\partial x_m}=A_m(x),
\ee
The matrix elements of $\tilde{A}(x)$ are $\mathbb{Q}$-linear combinations of logarithms. The arguments of the logarithms are known as \emph{letters}, while the set of linearly independent letters is known as \emph{alphabet}.
The main virtue of the canonical system of differential equations is that its solution is elementary, and it can be written for general $\epsilon$ in terms of a path-ordered exponential,
\begin{equation}
\label{eq:path-ordered}
f(x,\epsilon)= P e^{\epsilon \int_C d \tilde{A}} f(0,\epsilon)\,,
\end{equation}
where $P$ is the path ordering operator along the integration path $C$, connecting the boundary point to $x$, while $f(0,\epsilon)$ are boundary conditions for $f(x,\epsilon)$.
In practice it is convenient to express the solution as a power series around $\epsilon=0$. Denoting with $f^{(i)}(x)$ the coefficient of $\epsilon^i$, we have,
\be 
f(x)=\sum_i f^{(i)}(x) \epsilon^i,
\ee 
and the different orders of the solution are related by the following recursive relation,
\begin{equation}
\label{seriessolution}
 f^{(i)}(x)=\epsilon \int_C d\tilde{A}(x)  f^{(i-1)}+f^{(i)}(0).
\end{equation}
The previous relation shows that the solution is expressed to all orders of $\epsilon$ in terms of Chen iterated integrals~\cite{Chen-iterated}. The solution is a pure function of uniform weight corresponding to the order of the $\epsilon$ expansion.

The specific choice of the integral basis leading to the canonical form was achieved using the ideas outlined in~\cite{Henn:2013pwa}. In particular, it is expected that integrals with constant leading singularities~\cite{ArkaniHamed:2010gh} satisfy canonical differential equations. Using generalized cuts we look for combinations of integrals with simple leading singularities, that can be normalized to unity rescaling the candidate integrals. This typically leads to a form close to the canonical form. The remaining unwanted terms can be then algorithmically removed from the differential equations shifting the integral basis~\cite{Caron-Huot:2014lda,Gehrmann:2014bfa,Argeri:2014qva}.

\subsection{Polylogarithmic representation for algebraic alphabets}
\label{sec:solution canonical}

The alphabet (see appendix \ref{sec:appendix:alphabet} for the explicit alphabet of the integral families) of the canonical integrals discussed in this paper contains 8 independent square roots that cannot be simultaneously rationalized via a variable change. This means that it is not possible to integrate (\ref{seriessolution}) directly in terms of multiple polylogarithms~\cite{Goncharov:1998kja}.

However we can find an expression in terms of these functions by making a suitable ansatz in terms of polylogarithms of a given weight. The main task is to find suitable function arguments, as we discuss presently.
This strategy is streamlined using the concept of \emph{symbol}~\cite{Goncharov:1998kja,Brown:2009qja,Goncharov:2010jf} of an iterated integral. 
The symbol corresponds to the integration kernels defining the iterated integrals. Since the integral basis is chosen to be of uniform weight, the symbol of the solution is completely manifest in our differential equations approach. Denoting by $f_n^{(i)}$ the $n^{th}$ component of the basis at $\mathcal{O}(\epsilon^i)$, and by $\tilde{A}_{nm}$ the $n^{th}$-row, $m^{th}$-column entry of matrix $\tilde{A}$, we have the following expression for the symbol of $f_n^{(i)}$,
\begin{equation}
\label{symbolseriessolution}
\mathcal{S}( f^{(i)}_n(x)) = \sum_m \mathcal{S}( f^{(i-1)}_m(x)) \otimes \mathcal{S}(\tilde{A}_{nm}(x))\,.
\end{equation}
The corresponding polylogarithmic functions can be found proceeding in the following algorithmic steps (see also~\cite{Goncharov:2010jf,Duhr:2011zq}). First, one generates a list of function arguments as monomials in the letters appearing in the alphabet. For the classical polylogarithms ${\rm Li}_{n}(x)$, one requires that $1-x$ factorizes over
the letters appearing in the alphabet\footnote{When square roots are present it might be difficult to directly check factorization over the alphabet. In practice we can proceed as follows. We consider the logarithm of the function whose factorization we want to check, and we equate it to a generic linear combination of the logarithms of the alphabet (ansatz). Since additive constants are irrelevant at the symbol level, we derive the identity with respect to each variable. We then specialize the resulting system of equations for the (rational) free coefficients of the ansatz to many numeric values of the variables. If a solution exists the argument factorizes as desired over the alphabet and the solution defines the factorized form.} (a caveat is that in principle spurious letters might be needed~\cite{Duhr:2011zq}). For ${\rm Li}_{2,2}(x,y)$, the condition is that $1-x,1-y,1-x y$ factorize over the alphabet. Similar factorization properties are required for higher weight functions. Second, for each weight $i$, one chooses a maximal set of linearly independent functions for the alphabet. The linear independence can be verified using the symbol. By construction, we can solve the differential equations at every order in terms of this set of functions. Third, we determine the terms in the kernel of the symbol at weight $i$ by writing the most general ansatz in terms of the lower weight functions, and solving the differential equations at $\mathcal{O}(\epsilon^i)$ for the free coefficients of the ansatz. Finally, we recover transcendental additive constants imposing boundary conditions. Note that no assumptions were made on the rationality of the alphabet letters, so that the above steps generalize the algorithm of~\cite{Duhr:2011zq} to algebraic cases. Note also that, as opposed to a purely symbol-based approach, using the knowledge of the differential equations and of the boundary conditions, the solution is fully determined.

In practice the alphabet under consideration is quite large, and a reasonably fast computer implementation of the algorithm above up to weight four is challenging. We can nevertheless use the algorithm to reconstruct polylogarithmic functions up to weight two, for which the alphabet letters contributing to the result are a relatively small subset of the full alphabet. The full set of linearly independent dilogarithms for the four families is listed in appendix~\ref{sec:appendix:li2}.  

Having a representation of the weight two functions in terms of classical polylogarithms at hand is in fact very useful. As was shown in ref. \cite{Caron-Huot:2014lda}, this can be used to write down useful one-dimensional integral representations for the remaining weight-three and weight-four functions. 

Following  \cite{Caron-Huot:2014lda}, we use the Chen integral representation of the solution to write down a one-fold integral representation at weight three and four. Parametrizing the integration path $C$ with $\alpha\in [0,1]$, (\ref{seriessolution}) translates to an iterated integral,
\begin{equation}
f^{(i)}(x)=\epsilon \int_0^1 (\partial_{\alpha} \tilde{A}(\alpha)) f^{(i-1)}(\alpha) d\alpha+f^{(i)}(0)\,.
\end{equation}
In this language when the weight-two functions are known analytically, the weight-three functions are one-fold integrals. Initially, the weight-four functions are two-fold iterated integrals of differentials of logarithms, and they can be converted to one-fold integrals integrating by parts (see appendix \ref{sec:appendix:integral representation} for a detailed discussion).

The boundary conditions required to fix the solution are determined using the regularity of the pre-canonical integrals and the behavior of the algebraic factors defining the canonical basis in the boundary point. We find  it convenient to use the boundary point $x_1=x_2=x_3=0$.
The values of our integrals at this point correspond to the large heavy-quark limit
so that one can apply the corresponding well-known graph theoretical
prescriptions
~\cite{Smirnov:1990rz,Smirnov:1994tg,Smirnov:2002pj}.
In the limit all the canonical integrals vanish except those that factor into products of one-loop integrals of which one is massless and thus diverges in the limit. These are however known analytically to all orders~\cite{Gehrmann:1999as,Ellis:2007qk}.
 With this choice of the boundary point we can parametrize the integration path as,
\begin{equation}
\label{eq:contour}
x(\alpha)= \left\{ x_1\,\alpha,\,x_2\,\alpha,\,x_3\,\alpha \right\}\,,
\end{equation}
with $\alpha \in [0,1]$.

We have validated the analytic expressions performing numerical checks against the computer program FIESTA~\cite{Smirnov:2008py,Smirnov:2009pb,Smirnov:2013eza} for randomly selected points in the Euclidean region (\ref{eq:euclidean region}).

\section{Elliptic integral sectors}
\label{sec:elliptic sectors}

The last two integral sectors of Family $A$ (see appendix~\ref{sec:appendix:integral basis}),
integrals $f_{66}^A-f_{73}^A$, turn out to be expressed in terms of elliptic functions. Using the
language of the differential equations, the homogeneous part for sector
$I^{A}_{1,1,0,1,1,1,1,0,0}$ is not cast in canonical form, as the solution is expressed in
terms of complete elliptic integrals. In appendix \ref{sec:appendix:cut} we show that  these
properties can be verified a priori analyzing the maximal cut of the integrals. In
section~\ref{sec:general framework} we show that we can reduce the problem to the solution of
a second order differential equation. In section~\ref{sec:solution second order de} we
show that using a proper unidimensional parametrization of the integrals the relevant second
order differential equation can be solved with elementary techniques. In
section~\ref{sec:two-fold iterated integrals} we show that employing two auxiliary bases we
obtain a two-fold iterated integral representation of the integral sector.

The highest sector of Family $A$ is $I^{A}_{1,1,1,1,1,1,1,0,0}$. In this case the homogeneous
part of the differential equations can be cast in canonical form, however they depend via
inhomogeneous terms on the lower elliptic sector. In section~\ref{sec:highest sector} we write
the result as a three-fold integral. We found these integral representations to be suitable
for precise and reliable numerical evaluations. When implemented in Mathematica the evaluation
of both the elliptic sectors in one Euclidean point takes about 10 minutes using one
processor, with about eight-digit accuracy.

\bfig
\bc
\vspace*{5mm}
\[ 
\vcenter{
\hbox{
  \begin{picture}(0,0)(0,0)
\SetScale{0.65}
  \SetWidth{1}
\Line(-25,30)(-40,30)
\Line(25,30)(40,30)
\Line(-25,-30)(-40,-30)
\Line(25,30)(0,-30)
%
%
  \SetWidth{3.5}
\DashLine(25,-30)(40,-30){3}
\Line(-25,30)(-25,-30)
\Line(-25,-30)(25,-30)
\Line(25,-30)(25,30)
\Line(25,30)(-25,30)
%
%
%
%
\end{picture}}
}
\hspace{3.0cm}
\vcenter{
\hbox{
  \begin{picture}(0,0)(0,0)
\SetScale{0.65}
  \SetWidth{1}
\Line(-25,30)(-40,30)
\Line(25,30)(40,30)
\Line(-25,-30)(-40,-30)
\Line(25,30)(0,-30)
\CCirc(0,30){5}{0.9}{0.9}
  \SetWidth{3.5}
\DashLine(25,-30)(40,-30){3}
\Line(-25,30)(-25,-30)
\Line(-25,-30)(25,-30)
\Line(25,-30)(25,30)
\Line(25,30)(-25,30)
%
%
%
%
\end{picture}}
}
\hspace{3.0cm}
\vcenter{
\hbox{
  \begin{picture}(0,0)(0,0)
\SetScale{0.65}
  \SetWidth{1}
\Line(-25,30)(-40,30)
\Line(25,30)(40,30)
\Line(-25,-30)(-40,-30)
\Line(25,30)(0,-30)
\CCirc(-15,-30){5}{0.9}{0.9}
  \SetWidth{3.5}
\DashLine(25,-30)(40,-30){3}
\Line(-25,30)(-25,-30)
\Line(-25,-30)(25,-30)
\Line(25,-30)(25,30)
\Line(25,30)(-25,30)
%
%
%
%
\end{picture}}
}
\hspace{3.0cm}
\vcenter{
\hbox{
  \begin{picture}(0,0)(0,0)
\SetScale{0.65}
  \SetWidth{1}
\Line(-25,30)(-40,30)
\Line(25,30)(40,30)
\Line(-25,-30)(-40,-30)
\Line(25,30)(0,-30)
\CCirc(25,0){5}{0.9}{0.9}
  \SetWidth{3.5}
\DashLine(25,-30)(40,-30){3}
\Line(-25,30)(-25,-30)
\Line(-25,-30)(25,-30)
\Line(25,-30)(25,30)
\Line(25,30)(-25,30)
\Text(2,33)[c]{$(k_2\!+\!p_1)^2$}
%
%
\end{picture}}
}
\]
\vspace*{2mm}
\caption{\label{fig:lower elliptic}The four master integrals of the elliptic sector $I^{A}_{1,1,0,1,1,1,1,0,0}$.}
\ec
\efig
%

\bfig
\bc
\vspace*{8mm}
\[ 
\vcenter{
\hbox{
  \begin{picture}(0,0)(0,0)
\SetScale{0.65}
  \SetWidth{1}
\Line(-25,30)(-40,30)
\Line(25,30)(40,30)
\Line(-25,-30)(-40,-30)
\Line(0,-30)(0,30)
%
%
  \SetWidth{3.5}
\DashLine(25,-30)(40,-30){3}
\Line(-25,30)(-25,-30)
\Line(-25,-30)(25,-30)
\Line(25,-30)(25,30)
\Line(25,30)(-25,30)
%
%
%
%
\end{picture}}
}
\hspace{3.0cm}
\vcenter{
\hbox{
  \begin{picture}(0,0)(0,0)
\SetScale{0.65}
  \SetWidth{1}
\Line(-25,30)(-40,30)
\Line(25,30)(40,30)
\Line(-25,-30)(-40,-30)
\Line(0,-30)(0,30)
%
%
  \SetWidth{3.5}
\DashLine(25,-30)(40,-30){3}
\Line(-25,30)(-25,-30)
\Line(-25,-30)(25,-30)
\Line(25,-30)(25,30)
\Line(25,30)(-25,30)
\Text(0,33)[c]{$(k_2\!+\!p_1)^2$}
%
%
\end{picture}}
}
\hspace{3.0cm}
%
\vcenter{
\hbox{
  \begin{picture}(0,0)(0,0)
\SetScale{0.65}
  \SetWidth{1}
\Line(-25,30)(-40,30)
\Line(25,30)(40,30)
\Line(-25,-30)(-40,-30)
\Line(0,-30)(0,30)
%
%
  \SetWidth{3.5}
\DashLine(25,-30)(40,-30){3}
\Line(-25,30)(-25,-30)
\Line(-25,-30)(25,-30)
\Line(25,-30)(25,30)
\Line(25,30)(-25,30)
\Text(0,33)[c]{$(k_1\!-\!p_3)^2$}
%
%
\end{picture}}
}
\hspace{3.0cm}
\vcenter{
\hbox{
  \begin{picture}(0,0)(0,0)
\SetScale{0.65}
  \SetWidth{1}
\Line(-25,30)(-40,30)
\Line(25,30)(40,30)
\Line(-25,-30)(-40,-30)
\Line(0,-30)(0,30)
%
%
  \SetWidth{3.5}
\DashLine(25,-30)(40,-30){3}
\Line(-25,30)(-25,-30)
\Line(-25,-30)(25,-30)
\Line(25,-30)(25,30)
\Line(25,30)(-25,30)
\Text(0,33)[c]{$(k_2\!+\!p_1)^2(k_1\!-\!p_3)^2$}
%
%
\end{picture}}
}
\]
\vspace*{2mm}
\caption{\label{fig:higher elliptic}The four master integrals of the elliptic sector $I^{A}_{1,1,1,1,1,1,1,0,0}$.}
\ec
\efig
%


\subsection{Sector $I^{A}_{1,1,0,1,1,1,1,0,0}$}
\label{sec:general framework}
The integral sector $I^{A}_{1,1,0,1,1,1,1,0,0}$ has four master integrals, shown in fig.~\ref{fig:lower elliptic}, which are expressed in terms of elliptic functions, although its subtopologies do not involve them. We start by considering the following basis of \emph{finite} integrals,
\be
\begin{split} \label{eq:basis elliptic 4}
h_{1}(x,\ep)& =\epsilon ^4 (-x_1)^{3/2} I^{A}_{1,1,0,1,1,1,1,0,0}\,,\\ 
h_{2}(x,\ep)& =\epsilon^4 I^{A}_{2,1,0,1,1,1,1,0,0}\,,\\
h_{3}(x,\ep)& =\epsilon^3 I^{A}_{1,1,0,1,1,1,2,0,0}\,,\\
h_{4}(x,\ep)& =\epsilon^4 I^{A}_{1,1,0,1,1,1,1,0,-1}\,.
\end{split} 
\ee
%
We parametrize the integrals through the linear parametrization (\ref{eq:contour}),
and we define the differential equations with respect to the new parameter using the chain rule,
\begin{equation}
\partial_{\alpha} h(x(\alpha),\epsilon)= \sum_{i=1}^3 x_i \,\partial_{x_i} h(x(\alpha),\epsilon)\,,
\end{equation}
where $h$ is a vector, whose components are given in eq.~(\ref{eq:basis elliptic 4}).
The differential equations have the following form,
\begin{equation}
\label{eq:DE no expanded }
\partial_{\alpha} h(\alpha,\epsilon)= C^{(0)}(\alpha) h(\alpha,\epsilon)+\epsilon\, C^{(1)}(\alpha) h(\alpha,\epsilon)+\epsilon\, D^{(1)}(\alpha)\,g(\alpha,\epsilon)+\mathcal{O}(\epsilon^2)\,,
\end{equation}
where $g(\alpha,\epsilon)$ is the vector of the subtopologies, $C^{(0)}(\alpha)$ and
$C^{(1)}(\alpha)$ are $4\times 4$ matrices and $D^{(1)}(\alpha)$ is a $4\times 65$ matrix.
In particular, the matrix $C^{(0)}(\alpha)$ has the form,
\begin{equation}
\label{eq:de matrix ell}
C^{(0)}(\alpha)= 
 \begin{pmatrix}
  a_{1,1}  & a_{1,2} & 0 & 0 \\
  a_{2,1}  & a_{2,2} & 0 & 0 \\
  a_{3,1}  & a_{3,2} & a_{3,3}  & 0 \\
  a_{4,1}  & a_{4,2} & 0 &  a_{4,4} 
\end{pmatrix}.
\end{equation}
The last two integrals are decoupled from each other,  but this is not required for the applicability of the method described here. It is manifest that the equations for the first two integrals are coupled.  

We look for a solution in power series around $\epsilon=0$,
\begin{equation}
h(\alpha,\epsilon)=\sum_{i}  h^{(i)}(\alpha) \epsilon^i.
\end{equation}
The coefficients of the power series satisfy the following first order differential equations,
\begin{equation}
\label{eq:de elliptic expanded}
\partial_{\alpha} h^{(i)}(\alpha) = C^{(0)}(\alpha)h^{(i)}(\alpha)+\epsilon\, C^{(1)}(\alpha)h^{(i-1)}(\alpha)+\epsilon \,D^{(1)}(\alpha)\,g^{(i-1)}(\alpha)+\mathcal{O}(\epsilon^2)\, ,
\end{equation}
where $h^{(i)}(\alpha)$ is the unknown and the other terms define the inhomogeneous part. A two-by-two system of first order differential equations for the first two components of $h(\alpha)$ defines a second order differential equation for the first component,
\begin{equation}
\label{eq:second order DE}
\partial_{\alpha}^2 h_1^{(i)}(\alpha)+p_1(\alpha) \, \partial_{\alpha}h_1^{(i)}(\alpha)+q_1(\alpha)\, h_1^{(i)}(\alpha) = r_1^{(i)}(\alpha)\,,   
\end{equation}
where $p_1(\alpha)$ and $q_1(\alpha)$ depend on the matrix elements of $C^{(0)}(\alpha)$, and are the same for every $i$, while $r_1^{(i)}(\alpha)$ is a function of the inhomogeneous part of (\ref{eq:de elliptic expanded}). Once two homogeneous solutions of (\ref{eq:second order DE}), $y_1(\alpha)$ and $y_2(\alpha)$, have been found, a particular solution can be determined using the method of the variation of constants. In general we get,
\begin{equation}
\label{eq:sol elliptic 1}
h_1^{(i)}(\alpha)=c_1\, y_1(\alpha)+c_2 \,y_2(\alpha) -y_1 (\alpha)\int_0^{\alpha} dz  \frac{r_1^{(i)}(z )}{w(z )} y_2(z ) +y_2 (\alpha)\int_0^{\alpha}dz  \frac{r_1^{(i)}(z )}{w(z )} y_1(z )\,,
\end{equation}
where the arbitrary constants $c_i$ are fixed by the boundary conditions,
and where $w(\alpha)$ is the Wronskian of the homogeneous solutions,
\begin{equation}
w(\alpha)= y_2(\alpha)\, \partial_{\alpha}y_1(\alpha)-y_1(\alpha)\, \partial_{\alpha}y_2(\alpha)\,.
\end{equation}

Once $h_1^{(i)}(\alpha)$ is solved, we can determine the remaining components of $h^{(i)}(\alpha)$. From (\ref{eq:de matrix ell}) it follows that $h_2^{(i)}(\alpha)$ can be obtained from $h_1^{(i)}(\alpha)$ and its first derivative. In this way the expression of $h_2^{(i)}(\alpha)$ involves the same number of repeated integrations as $h_1^{(i)}(\alpha)$. In order to solve the last two integrals we solve the respective first order differential equations, which depend on $h_1^{(i)}(\alpha)$ and $h_2^{(i)}(\alpha)$ via the inhomogeneous terms. This shows that, when computed in this way, $h_3^{(i)}(\alpha)$ and $h_4^{(i)}(\alpha)$ involve one more repeated integration than $h_1^{(i)}(\alpha)$ and $h_2^{(i)}(\alpha)$.
In order to optimize the numerical evaluation it is important to get rid of the extra integration. Furthermore, since at $\mathcal{O}(\epsilon^4)$
these integrals would be expressed in terms of five iterated integrations, one integration must be spurious. 
In the non-elliptic case one is able to remove extra integrations using integration by parts. However in the elliptic case in order to perform an integration by parts one needs to integrate over elliptic functions, which is in general not possible analytically. 
We show how this is done in section~\ref{sec:two-fold iterated integrals}.


\subsection{Solution of the second order differential equation}
\label{sec:solution second order de} 
The possibility of solving algorithmically a second order differential equation is related to the number of its singular points, including the point at infinity. If there are up to three singular points the equation can be cast in the form of the hypergeometric equation and two linearly independent solutions can be expressed in terms of hypergeometric functions~\cite{watson}. Similar algorithms exist when four singular points are present. On the other hand if more than four singular points are present the solution requires a case by case analysis.

After differentiating with respect to the Mandelstam variables, the second order differential equation for $I_{1,1,0,1,1,1,1,0,0}^{A}$ has six singular points. We show that using the parametrization (\ref{eq:contour}) the solution can be reduced to the three singular point case. 

Once $h_1(x(\alpha),\epsilon)$ is made explicit as in (\ref{eq:basis elliptic 4}), the coefficients of the second order differential equation (\ref{eq:second order DE}) are,
\be
p_1(\alpha)=\frac{2 x_1 \left(\alpha \, x_1 \left(x_2-x_3\right){}^2-4 \left(x_2 \left(x_1-x_3\right)+x_3 \left(x_1+x_3\right)\right)\right)}{ d_1(\alpha)}\,,
\ee
and,
\be
q_1(\alpha)=\frac{x_1^2 \left(x_2-x_3\right){}^2}{4 d_1(\alpha)}\,,
\ee 
where,
\begin{equation}
d_1(\alpha)=x_1^2\, \alpha^2 \left(x_2-x_3\right){}^2-8 x_1\, \alpha \left(x_2 (x_1-x_3)+x_3 (x_1+x_3)\right)+16 (x_1+x_3)^2\,.
\end{equation}
We see that after using parametrization (\ref{eq:contour}) we are left with three singular points, which are the two roots of $d_1(\alpha)=0$ and the point at infinity.
The homogeneous solutions of (\ref{eq:second order DE}) can be then readily found\footnote{We have found the Mathematica built-in function DSolve to be adequate. In alternative it is possible to use the algorithm of~\cite{watson}.} to be
\begin{equation}
\label{solf2homo}
y_1(\alpha)= K\left(\frac{1}{2}-\frac{k(\alpha)}{2}\right),\qquad y_2(\alpha)=K\left(\frac{1}{2}+\frac{k(\alpha)}{2}\right),
\end{equation} 
where the function $k(z )$ is,
\begin{equation}
k(z )= \frac{ \left(x_2-x_3\right){}^2\, x_1\, z  -4 \left(x_2 (x_1-x_3)+x_3 (x_1+x_3)\right)}{8 \sqrt{x_1\, x_3\, x_2 \, (x_1+x_3-x_2)}}\,,
\end{equation}
and $K(z)$ is the complete elliptic integral of the first kind\footnote{Note that also a different convention exists for the definition of complete elliptic integrals such that, compared to our definition, the argument is replaced by its squared at the level of the integrand.},
\begin{equation}
K(z )=\int_0^1 \frac{dt}{\sqrt{(1-t^2)(1-z \, t^2)}}\,.
\end{equation}
The complete elliptic integral of the second kind is defined as,
\be
E(z )=\int_0^1 \frac{\sqrt{1-z \, t^2}}{\sqrt{1-t^2}}dt\,.
\ee
We have the following relations for the derivatives of the complete elliptic integrals,
\be 
\label{eq:de k}
\frac{ d K(z )}{dz }=\frac{E(z  )-(1-z  ) K(z  )}{2 (1-z  ) z }\,,
\ee
and,
\be
\frac{d E(z )}{dz }=\frac{E(z  )-K(z  )}{2 z  }\,.
\ee
Since $h_2^{(i)}(\alpha)$ is a linear combination of $h_1^{(i)}(\alpha)$ and its first derivative, it is expressed in terms of complete elliptic integrals of the first and second kind, of the same arguments as in (\ref{solf2homo}). 
The Wronskian of the two homogeneous solutions is defined in terms of the derivatives above. Its expression is a rational function of the integration variable $\alpha$, and in our case it reads,
\begin{equation}
\label{eq:expr wronskian}
w(\alpha)=\frac{4 \pi  x_1 \sqrt{x_1\, x_3\, x_2  \left(x_1+x_3-x_2\right)}}{d_1(\alpha)}.
\end{equation}
This property can be proven by using the Legendre identity,
\be
E(z )K(1-z )+E(1-z )K(z )-K(z )K(1-z )=\frac{\pi}{2}\,.
\ee

Thanks to the overall normalization factor we chose for $h_1(x,\epsilon)$, it is elementary to determine boundary conditions and use them to fix the free constants of the general solution~(\ref{eq:sol elliptic 1}). Integral $I^{A}_{1,1,0,1,1,1,1,0,0}$ is regular for $\alpha=0$, so that $h_1(0,\epsilon)=\partial_\alpha h_1(0,\epsilon)=0$ and $c_1=c_2=0$.


\subsection{Auxiliary bases and solution in terms of two-fold iterated integrals}
\label{sec:two-fold iterated integrals}

Since we need to evaluate the components of $h$ (\ref{eq:basis elliptic 4}) through $\mathcal{O}(\epsilon^4)$, all the $I$ integrals of 
eq.~(\ref{eq:basis elliptic 4}) need to be computed through $\mathcal{O}(\epsilon^0)$, except $I^{A}_{1, 1, 0, 1, 1, 1, 2, 0, 0}$ which must be evaluated
through $\mathcal{O}(\epsilon)$. Higher orders are irrelevant for two-loop processes. In general, the result for a master integral at $\mathcal{O}(\epsilon^i)$ is obtained integrating over subtopologies through $\mathcal{O}(\epsilon^{i-1})$ and, if coupled to them, over integrals of the same topology at $\mathcal{O}(\epsilon^i)$. In section~\ref{sec:solution canonical} we saw that weight-two functions can be expressed in terms of logarithms and dilogarithms, and weight-three functions can be reduced to one-fold integrals. This implies that, because of the general form of (\ref{eq:DE no expanded }), $h_{3}^{(3)}(\alpha)$ is expressed in terms of one-fold integrals, while $h_{1}^{(4)}(\alpha)$ and $h_{2}^{(4)}(\alpha)$ are expressed in terms of up to two-fold integrals. On the other hand $h_3^{(4)}(\alpha)$ and $h_4^{(4)}(\alpha)$ involve three-fold iterated integrals. 

In order to avoid considering more than two iterated integrations we introduce two auxiliary bases. We look for bases with differential equations of the form of (\ref{eq:DE no expanded }), where the first integral is $h_1(\alpha,\epsilon)$ and the second integral is linearly independent of $h_1(\alpha,\epsilon)$ and $h_2(\alpha,\epsilon)$, the auxiliary integrals being independent of each other. In this way we compute the two remaining integrals as linear combinations of $h_1(\alpha,\epsilon)$ and its first derivative, generating at most two-fold iterated integrals. In practice we found two auxiliary bases equal to basis (\ref{eq:basis elliptic 4}) modulo replacing, in turn, $h_2(\alpha,\epsilon)$ with $\epsilon^4 I^{A}_{1, 2, 0, 1, 1, 1, 1, 0, 0}$ and $\epsilon^4 I^{A}_{1, 1, 0, 1, 1, 1, 2, -1, 0}$.
The full (finite) basis for the integral sector is then chosen to be,
\be
\begin{split}\label{eq:full basis elliptic 4}
f_{66}^A &=\epsilon^4 (-x_1)^{3/2} I^{A}_{1, 1, 0, 1, 1, 1, 1, 0, 0}\,,\\
f_{67}^A &=\epsilon^4 (-x_1)^{3/2} \, x_1 I^{A}_{2, 1, 0, 1, 1, 1, 1,0, 0}\,,\\
f_{68}^A &=\epsilon^4 (-x_1)^{3/2}\, x_1 I^{A}_{1, 2, 0, 1, 1, 1, 1, 0, 0}\,,\\
f_{69}^A &=\epsilon^4 (-x_1)^{3/2} I^{A}_{1, 1, 0, 1, 1, 1, 2, -1, 0}\,.
\end{split}
\ee 

Interestingly, if we consider the differential equations for $f^A_{66}-f^A_{69}$, they are fully coupled and cannot be solved directly. We could nevertheless solve them with the help of auxiliary bases.


\subsection{Sector $I^{A}_{1,1,1,1,1,1,1,0,0}$} 
\label{sec:highest sector}
The highest elliptic sector is $I^{A}_{1,1,1,1,1,1,1,0,0}$. It has four master integrals, shown in fig.~\ref{fig:higher elliptic}, and it depends on the elliptic subsector $I^{A}_{1,1,0,1,1,1,1,0,0}$ via inhomogeneous terms in the differential equations. Using the criteria outlined in~\cite{Henn:2013pwa} we can find a basis satisfying,
\begin{equation}
\label{eq:de highest}
\partial_{\alpha} v (\alpha,\epsilon)= \epsilon \,F^{(1)}(\alpha) v(\alpha,\epsilon)+G^{(0)}(\alpha)g(\alpha,\epsilon)+\epsilon \,G^{(1)}(\alpha)\,g(\alpha,\epsilon)+\mathcal{O}(\epsilon^2)\,.
\end{equation}
$v(\alpha,\epsilon)$ is a four-dimensional basis vector for the highest elliptic sector, $g(\alpha,\epsilon)$ is the vector of the subtopologies, $F^{(1)}(\alpha)$ is a $4\times 4$ matrix, $G^{(0)}(\alpha)$ and $G^{(1)}(\alpha)$ are $4\times 69$ matrices.
The homogeneous part is in canonical form, while this is not the case for the subtopologies. When solving the above equation for a given power of $\epsilon$, we have to integrate over subsectors of the same order due to the $G^{(0)}(\alpha)$ matrix.
For numerical optimization it is convenient to get rid of such integrals. Matrix elements of $G^{(0)}(\alpha)$ corresponding to non-elliptic subsectors are removed with a basis shift, as described in~\cite{Caron-Huot:2014lda,Gehrmann:2014bfa}. In order to remove $G^{(0)}(\alpha)$ entries corresponding to elliptic subsectors we proceed as follows. Let us consider the $i^{th}$ component of $v$, which fulfills the equation,
\begin{equation}
\partial_{\alpha} v_i(\alpha,\epsilon)= \sum_{j=1}^2 k_{ij}(\alpha)e_j(\alpha,\epsilon)+\mathcal{O}(\epsilon)\,,
\end{equation}
where $k_{ij}(\alpha)$, with $i=1,\ldots,4$ and $j=1,2$, are known algebraic functions and $e_1$, $e_2$ are two coupled integrals of an elliptic subsector, satisfying,
\begin{equation}
\label{eq:de e highest sector}
\partial_{\alpha} e_i(\alpha,\epsilon)= \sum_{j=1}^2 a_{ij}(\alpha) e_j(\alpha)+\mathcal{O}(\epsilon)\,.
\end{equation}
 We shift $v_i(\alpha,\epsilon)$ according to,
\begin{equation}
\label{eq:de elliptic coupled}
v_i(\alpha,\epsilon)\rightarrow v_i(\alpha,\epsilon)+\sum_{j=1}^2 b_{ij}(\alpha)e_j(\alpha,\epsilon)\,,
\end{equation}
where $b_{ij}(\alpha)$ are functions to be determined. After the basis shift the equation for $v_i$ reads,
\begin{equation}
\partial_{\alpha} v_i(\alpha,\epsilon)= \sum_{j=1}^2 \big(\partial_{\alpha}b_{ij}(\alpha)+\sum_{k=1}^2 a_{kj}(\alpha)b_{ik}(\alpha)+k_{ij}(\alpha)\big)e_{j}(\alpha)+\mathcal{O}(\epsilon)\,.
\end{equation}
In order to remove  terms proportional to $e_1(\alpha,\epsilon)$ and $e_2(\alpha,\epsilon)$, 
their coefficients must vanish, \emph{i.e.} $b_{ij}(\alpha)$ must fulfill the equations,
\begin{equation}
\label{eq:de shift}
\partial_{\alpha} b_{ij}(\alpha)= -\sum_{k=1}^2 a_{kj}(\alpha) b_{ik}(\alpha)-k_{ij}(\alpha)\,,
\end{equation}
with $j=1,2$.
For fixed $i$,  the above equation is a two-by-two system of first order differential equations. The matrix defining the system is the transpose of the matrix defining  (\ref{eq:de e highest sector}). This implies that if $y_1(\alpha)$ and  $y_2(\alpha)$ are the homogeneous solutions of (\ref{eq:de e highest sector}) and $w(\alpha)$ is their Wronskian, the solutions of (\ref{eq:de shift}) are,
\be
\label{eq:sol transpose}
c \,\frac{y_1(\alpha)}{w(\alpha)}\,,\qquad c\, \frac{y_2(\alpha)}{w(\alpha)}\,,
\ee
where $c$ is an overall constant. Their Wronskian is $c^2/w(\alpha)$.  Therefore with the method of the variation of constants the full expression for $b_{i1}(\alpha)$ reads,
\begin{equation}
\label{eq:sol elliptic shift}
b_{i1}(\alpha)= -\frac{y_1(\alpha)}{w(\alpha)} \int_1^{\alpha} d t\, L_i(t)\,y_2(t)+\frac{y_2 (\alpha)}{w(\alpha)}\int_1^{\alpha}d t \,L_i(t)  y_1(t)\,,
\end{equation}
where  $L_i(\alpha)$ are functions of $ k_{i1}(\alpha)$ and $k_{i2}(\alpha)$, and where two arbitrary integration constants have been set to zero. In addition, we set the lower integration bound to $1$ but we have the freedom to choose a different value. Usually this is dictated by the properties of the integrand, that might have non-integrable singularities for specific integration bounds. Once $b_{i1}(\alpha)$ is known it is elementary to obtain $b_{i2}(\alpha)$ using the same differential equations.

For sector $I^{A}_{1,1,1,1,1,1,1,0,0}$ the integrals that need to be shifted are $f_{71}^A$ and $f_{73}^A$, as $e_1$ and $e_2$  defined via (\ref{eq:de elliptic coupled}) are equal to $f_{66}^A$ and $f_{67}^A$ respectively. $y_1(\alpha)$ and $y_2(\alpha)$ are the same as those of (\ref{solf2homo}) and,
\be
\label{eq:shift integrands}
 L_2(z )=\frac{x_1(x_1-x_2)}{(4-x_1 \,z )^{3/2} }\,,\qquad L_4(z )=\frac{x_1(x_1+x_3)}{(-x_1 \,z )^{3/2}}\,,
\ee 
while $L_1$ and $L_3$ vanish.

In general the integrals of (\ref{eq:sol elliptic shift}) are not known analytically in closed form. Since after the basis shift they will contribute to the matrix elements of the differential equations, one might wonder if such a basis change is convenient in practice, as our main goal was to get rid of one integration. In practice, because of the simple form of (\ref{eq:shift integrands}), its numerical evaluation takes $\mathcal{O}(10^{-3})$ sec. 
In this form the result for the elliptic sector at $\mathcal{O}(\epsilon^4)$ is in terms of three-fold integrals, while their numerical performance is comparable to the one of two-fold integrals. Alternatively, it is possible to series expand the complete elliptic integrals of eq. (\ref{eq:sol elliptic shift}) and then perform the integrations analytically\footnote{We series expand the complete elliptic integrals using well known results. The expansion around a generic point $z_0$ will involve powers of $z-z_0$, and factors of $\log(z-z_0)$ if $z_0$ is a singular point. It is then possible to perform the integrations analytically when considering the elementary functions of eq. (\ref{eq:shift integrands}).}. In this way the result for the integral sector can be expressed in terms of two-fold integrals\footnote{In order to get rid of the extra integration, one could have performed an integration by parts after solving directly eq.~(\ref{eq:de highest}). Also this method introduces integrals over complete elliptic integrals and algebraic functions in the integrands of the solution. However such integrals are not as simple as the ones introduced by the basis shift, and the integration over the series expanded complete elliptic integrals is not straightforward.}.

\section{The class of functions}
\label{sec:symbols}
In order to discuss the general structure of the solution of sector $I_{1,1,0,1,1,1,1,0,0}^{A}$ let us introduce the following shorthands for the complete elliptic integrals defined in section \ref{sec:solution second order de},
\be 
\begin{split}
K^{(1)} (\alpha)=K\left(\frac{1}{2}+\frac{k(\alpha)}{2}\right)\,,\qquad K^{(-1)}(\alpha)=E\left(\frac{1}{2}-\frac{k(\alpha)}{2}\right),\\
E^{(1)}(\alpha)=E\left(\frac{1}{2}+\frac{k(\alpha)}{2}\right)\,,\qquad E^{(-1)}(\alpha)=E\left(\frac{1}{2}-\frac{k(\alpha)}{2}\right).\\
\end{split}
\ee
Integrals $f_{66}^{A,(4)}-f_{69}^{A,(4)}$ are expressed as linear combinations of the class of functions,
\be
\label{eq:class lower ell}
\mathcal{E}^{(\sigma)}(1)\int_0^1 \mathcal{F}(t) \mathcal{E}^{(-\sigma)}(t) dt,\,\\
\ee
where $\mathcal{E}^{(\sigma)}$ can be one of the following complete elliptic integrals,
\be 
K^{(\sigma)}(\alpha),\qquad E^{(\sigma)}(\alpha),
\ee
where $\sigma\in\{-1,1\}$. $\mathcal{F}(t)$ denotes a linear combination of pure weight-two and weight-three functions, belonging to the subtopologies, multiplied by either derivatives of logarithms or derivatives of algebraic functions, with respect to $\alpha$~\footnote{In a few cases also algebraic functions that are derivatives of (combinations of) incomplete elliptic integrals appear. However this result requires further investigation as a reparametrization of the square roots might reduce them to derivatives of algebraic or logarithmic functions.}. Interestingly, weight-three functions are never multiplied by derivatives of logarithms, but only by the following simple inverse square roots (modulo functions depending only on rescaled Mandelstam invariants),
\be
\frac{1}{\sqrt{\alpha}},\qquad\frac{1}{\sqrt{4-x_1 \alpha}}.
\ee
The same class of functions has been found in~\cite{Tancredi:2016talk} for the massive crossed triangle. See~\cite{Bloch:2013tra,Bloch:2014qca,Broedel:2014vla,Bloch:2016izu} for results in terms of elliptic polylogarithms~\cite{2011arXiv1110.6917B}, and~\cite{Adams:2014vja,Adams:2015gva,Adams:2015ydq,Adams:2016xah} for a related class of functions. 

In order to decouple integral sector $I_{1,1,1,1,1,1,1,0,0}^A$ from sector $I_{1,1,0,1,1,1,1,0,0}^A$, in section \ref{sec:highest sector} we  performed a non-algebraic basis shift of $f_{71}^A$ and $f_{73}^A$, involving integrals of complete elliptic integrals, that we denote here with the following shorthands,
\be
\begin{split}
\tilde{K}_i^{(1)}(\alpha)=\int_1^{\alpha} L_{i}(t) K^{(1)}(t)dt,\qquad\tilde{K}_i^{(-1)}(\alpha)=\int_1^{\alpha} L_{i}(t) K^{(-1)}(t)dt,
\end{split}
\ee 
where $L_i(t)$ are those of eq.~(\ref{eq:shift integrands}).
For this reason the result for the highest elliptic sector is not directly expressed in terms of iterated integrals of the form of eq.~(\ref{eq:class lower ell}), though such expressions can be immediately obtained by solving the differential equations without performing the non-algebraic basis shift. Integrals $f_{70}^{A,(4)}-f_{73}^{A,(4)}$ are linear combinations of polylogarithmic functions and of the class of functions,
\be
\int_0^1 \mathcal{G}(t) \mathcal{E}^{(\sigma)}(t) \tilde{K}_i^{(-\sigma)}(t)dt \,.
\ee
$\mathcal{G}(t)$ has the same properties as $\mathcal{F}(t)$ described above, but the prefactors of pure weight-three functions are any of the
algebraic functions,
\be
\frac{1}{\sqrt{\alpha}},\qquad\frac{1}{\sqrt{4-x_1 \alpha}},\qquad\sqrt{\alpha}.
\ee

\section{Conclusion and perspectives}
\label{sec:conclusions}
In this paper we presented the analytic computation of all the planar master integrals which are necessary to evaluate the two-loop amplitudes
for Higgs~$\to 3$~partons, with the full heavy-quark mass dependence. They occur in the NNLO corrections to fully inclusive Higgs production
and in the NLO corrections to Higgs plus one jet production in hadron collisions. The result is expressed in terms of iterated integrals over both algebraic and elliptic kernels. This is the first time that Feynman integrals for four-point multiscale amplitudes involving elliptic functions are computed in a fully analytic way. While it was generally believed that the analytic computation of multiscale loop integrals with many internal massive lines was out of reach with present analytic tools, this work shows that new ideas involving the proper parametrization of the integrals, an optimal basis choice, and the subsequent solution with the differential equations method in terms of elliptic iterated integrals, are effective to treat such problems.

The computation of the non-elliptic integral sectors has been performed with the differential equations method applied to a set of basis integrals defined to be pure functions of uniform weight. The presence of many square roots that cannot be simultaneously rationalized makes the direct solution of these equations in terms of multiple polylogarithms not possible. We have shown that the Chen iterated integral representation plus the knowledge of the boundary conditions provide the information needed to integrate the system in terms of a minimal polylogarithmic basis, circumventing in this way the necessity to rationalize the square roots of the alphabet. To do so we used an algorithm for the integration of symbols with general algebraic alphabets, generalizing well established algorithms for the rational case.

We have seen that the crucial point for the computation of the elliptic sectors is the solution of the associated homogeneous second order differential equation. We noticed that a very simple univariate reparametrization of the integrals makes the equation elementary and standard tools are sufficient to solve it. The central point is that the fewer singular points are present in higher-order differential equations, the simpler is their solution. It will be important to further investigate and develop the idea of what is the proper parametrization of the integrals yielding the simplest singular structure of the equations. The univariate parametrization has also the benefit that only one set of differential equations has to be solved, while in the traditional approach one has to iteratively solve multiple sets of equations, one for each variable, which might be highly non-trivial when elliptic functions are involved.

In contrast to the non-elliptic sectors, we did not use the notion of canonical basis for the elliptic sectors. Instead, we showed that the problem can be completely solved in total generality, once the relevant higher order homogeneous equations have been solved. However it will be important to extend the notion of canonical basis to elliptic cases. First, this will clarify the class of functions needed to represent the answer -- in our case we used a rather general class that might still contain spurious information. Second, it is natural to expect that the explicit results for canonical integrals will be relatively compact. In order to define a canonical basis in the elliptic case, the notion of leading singularity has to be generalized, which is beyond the scope of the present paper (see appendix \ref{sec:appendix:cut} for a discussion about the maximal cut of those integrals, which would be the starting point for defining a generalization of leading singularity in the elliptic case). In particular, we know \cite{Henn:2014qga1,Henn:2014qga,Lee:2014ioa} that it is possible to obtain a form of the differential equations with only Fuchsian singularities and linear in $\epsilon$. This is valid for any Feynman integral and it is another natural starting point for finding a canonical basis.

We showed that for the sake of stable and precise numerical evaluations we can express elliptic iterated integrals through $\mathcal{O}(\epsilon^4)$
in terms of one and two-fold iterated integrals for the non-elliptic and elliptic sectors, respectively. We found these representations suitable for numerical evaluation. In principle, as the integrands are known functions, it should be possible to achieve a series representation of the solution, though we did not attempt it as the integral representation already showed satisfying performance. It will be important to develop general purpose numerical routines for elliptic iterated integrals, so that one can take advantage of such analytic expressions also when higher loop orders are considered, \emph{i.e.} when more iterated integrals are needed.

\acknowledgments

We would like to thank Claude Duhr and Yang Zhang for useful discussions and for reading parts of the manuscript. Part of the algebraic manipulations required in this work were carried out with {\tt
FORM}~\cite{Kuipers:2012rf}. The Feynman diagrams were drawn with {\tt
Axodraw}~\cite{Vermaseren:1994je}.  VDD and HF were partly supported by the Research Executive
Agency (REA) of the European Union, through the Initial Training Network LHCPhenoNet under
contract PITN-GA-2010-264564. HF is supported through the Initial Training Network HiggsTools
under contract PITN-GA-2012-316704.  JMH is supported in part by a GFK fellowship and by the
PRISMA cluster of excellence at Mainz university.

\appendix

\section{Integral basis}
\label{sec:appendix:integral basis}

In this appendix, we provide the explicit form of the integral families we used to parametrize the integrals defined in eq.~(\ref{Integrals}). We call them: family $A$, $B$, $C$, and $D$.

For each family, we perform an independent reduction to the master integrals. Then we perform a change of basis that maps the master integrals into the canonical form. We give such a canonical basis for each family
separately. 
The canonical master integrals are labeled with $f^{i}_{n}$, with $i\in\{A,B,C,D\}$ and $n=1,...,N$, where $N$ is the number of master integrals of the family under consideration.
The elliptic sectors correspond to eight integrals of family $A$, labeled with $f^A_{66}$--$f^A_{73}$. These integrals are not in canonical form, as discussed in section~\ref{sec:elliptic sectors}.

For each family of integrals we define the corresponding system of differential equations, that we then solve as discussed in sections \ref{sec:DE} and \ref{sec:elliptic sectors}. 

Note that, in general, there is an overlap among the master integrals of the different families.
%
Making the appropriate correspondences, we can reduce the process to the computation of 125 master integrals.
In the next appendix, we draw these 125 (pre-canonical) master integrals and we link them to the corresponding canonical form.

We label with $p_1$, $p_2$, and $p_3$ the momenta of the massless partons, and with $p_4 = p_1+p_2+p_3$ the momentum of the Higgs. The loop momenta are labeled with $k_1$ and $k_2$. Finally, we use the shorthand $p_{ij}=p_i+p_j$.

Family $A$ is defined by the nine propagators,
\begin{gather}
 d_1^{A} = m^2-k_1^2, \quad d_2^{A}  = m^2-(k_1+p_{12})^2, \quad d_3^{A} = m^2-k_2^2,\nn\quad\quad\\\label{eq:Family A}
 d_4^{A} = m^2-(k_2+p_{12})^2, \quad d_5^{A} = m^2-(k_1+p_{1})^2, \quad d_6^{A} = -(k_1-k_2)^2,\\
 d_7^{A} = m^2-(k_2-p_3)^2, \quad d_8^{A} = -(k_2+p_1)^2, \quad d_9^{A} = -(k_1-p_3)^2,\quad\quad\nn
\end{gather}
with the extra restriction that $a_8$ and $a_9$ are non-positive. The family contains 73 master integrals. Below, we give the basis transformation between pre-canonical and canonical forms.
\begin{align}
f^A_{1} &= \ep^2 I^{A}_{0,0,0,0,2,0,2,0,0}\,,\nn\\
f^A_{2} &= \ep^2 {x_2} I^{A}_{0,2,0,0,0,1,2,0,0}\,,\nn\\
f^A_{3} &= \ep^2 \sqrt{4 {}-{x_2}} \sqrt{-{x_2}} \Big( I^{A}_{0,2,0,0,0,1,2,0,0}/2 + I^{A}_{0,2,0,0,0,2,1,0,0} \Big)\,,\nn\\
f^A_{4} &= \ep^2 x_1 I^{A}_{0,2,2,0,0,1,0,0,0}\,,\nn\\
f^A_{5} &= \ep^2 \sqrt{4 {}-x_1} \sqrt{-x_1} \Big( I^{A}_{0,2,2,0,0,1,0,0,0}/2 + I^{A}_{0,2,1,0,0,2,0,0,0} \Big) \,,\nn\\
f^A_{6} &= \ep^2 \sqrt{4 {}-x_1} \sqrt{-x_1} I^{A}_{0,0,2,1,2,0,0,0,0}\,,\nn\\
f^A_{7} &= \ep^2 \sqrt{4 {}-{x_2}} \sqrt{-{x_2}} I^{A}_{0,0,0,2,2,0,1,0,0}\,,\nn\\
f^A_{8} &= \ep^3 ({x_2}-x_1) I^{A}_{1,1,0,0,0,1,2,0,0}\,,\nn\\
f^A_{9} &= \ep^2 {} ({x_2}-x_1) I^{A}_{1,1,0,0,0,1,3,0,0}\,,\nn\\
f^A_{10} &= -\ep^2 \frac{\sqrt{4 {}-x_1}}{4 \sqrt{-x_1}} \Big( 2 \ep ({x_2}+x_1) I^{A}_{1,1,0,0,0,1,2,0,0} - 4 {} ({x_2}+x_1) I^{A}_{1,1,0,0,0,1,3,0,0} \nn \\
& \;\;\;\; + 4 x_1 I^{A}_{2,1,0,0,0,1,2,0,-1} + {x_2} I^{A}_{0,2,0,0,0,1,2,0,0} \Big)\,,\nn\\
f^A_{11} &= \ep^2 x_3 I^{A}_{0,0,0,0,2,1,2,0,0}\,,\nn\\
f^A_{12} &= \ep^2 \sqrt{4 {}-x_3} \sqrt{-x_3} \Big( I^{A}_{0,0,0,0,2,1,2,0,0} / 2 + I^{A}_{0,0,0,0,2,2,1,0,0} \Big)\,,\nn\\
f^A_{13} &= \ep^3 ({x_2}-x_1) I^{A}_{2,0,0,1,0,1,1,0,0}\,,\nn\\
f^A_{14} &= \ep^2 {} ({x_2}-x_1) I^{A}_{3,0,0,1,0,1,1,0,0}\,,\nn\\
f^A_{15} &= \ep^2 \frac{\sqrt{4 {}-{x_2}}}{\sqrt{-{x_2}} (2  - x_1)} \bigg( \ep \frac{2 {} {x_2} - x_1 (x_1-{x_2})}{2} I^{A}_{2,0,0,1,0,1,1,0,0} +  x_1 (x_1-x_2) I^{A}_{3,0,0,1,0,1,1,0,0} \nn \\
& \;\;\;\; + \frac{{x_2} \big( {} {x_2} + x_1 (x_1-{x_2}) \big)}{x_1-{x_2}} I^{A}_{2,0,-1,2,0,1,1,0,0} - \frac{x_1 \big( 4 {} {x_2}+x_1 (x_1-{x_2}) \big)}{4 \, (x_1-{x_2}) } I^{A}_{0,2,2,0,0,1,0,0,0} \bigg) \nn \\
f^A_{16} &= \ep^3 x_3 I^{A}_{1,0,0,0,1,1,2,0,0}\,,\nn\\
f^A_{17} &= \ep^3 ({x_2}-x_1) I^{A}_{0,2,1,0,0,1,1,0,0}\,,\nn\\
f^A_{18} &= \ep^3 x_1 I^{A}_{0,1,2,0,1,1,0,0,0}\,,\nn\\
f^A_{19} &= \ep^3 ({x_2}-x_3) I^{A}_{0,1,0,0,1,1,2,0,0}\,,\nn\\
f^A_{20} &= \ep^3 x_1 I^{A}_{0,0,1,1,2,1,0,0,0}\,,\nn\\
f^A_{21} &= \ep^2 {} x_1 I^{A}_{0,0,1,1,3,1,0,0,0}\,,\nn\\
f^A_{22} &= \ep^2 \sqrt{4 {}-x_1} \sqrt{-x_1} \Big( \ep I^{A}_{0,0,1,1,2,1,0,0,0}/2 - {} I^{A}_{0,0,1,1,3,1,0,0,0} + I^{A}_{0,0,2,1,2,1,0,-1,0} \Big)\,,\nn\\
f^A_{23} &= \ep^3 ({x_2}-x_3) I^{A}_{0,0,0,1,2,1,1,0,0}\,,\nn\\
f^A_{24} &= \ep^2 {} ({x_2}-x_3) I^{A}_{0,0,0,1,3,1,1,0,0}\,,\nn\\
f^A_{25} &= - \ep^2 \frac{\sqrt{4 {}-{x_2}}}{4 \sqrt{-{x_2}}} \Big( 2 \ep ({x_2}+x_3) I^{A}_{0,0,0,1,2,1,1,0,0} - 4  ({x_2}+x_3) I^{A}_{0,0,0,1,3,1,1,0,0} \nn \\
& \;\;\;\; + 4 {x_2} I^{A}_{0,0,0,2,2,1,1,-1,0} + x_3 I^{A}_{0,0,0,0,2,1,2,0,0} \Big) \nn \\
f^A_{26} &= \ep^3 ({x_2}-x_1) I^{A}_{0,0,1,1,2,0,1,0,0}\,,\nn\\
f^A_{27} &= \ep^2 (4 {}-x_1) x_1 I^{A}_{2,1,2,1,0,0,0,0,0}\,,\nn\\
f^A_{28} &= \ep^2 \sqrt{4 {}-{x_2}} \sqrt{-{x_2}} \sqrt{4 {}-x_1} \sqrt{-x_1} I^{A}_{2,1,0,2,0,0,1,0,0}\,,\nn\\
f^A_{29} &= \ep^3 \sqrt{4 {}-x_1} \sqrt{-x_1} x_1 I^{A}_{1,1,2,1,1,0,0,0,0}\,,\nn\\
f^A_{30} &= \ep^3 x_1 I^{A}_{1,1,0,0,1,0,2,0,0}\,,\nn\\
f^A_{31} &= \ep^3 \sqrt{4 {}-x_1} \, \sqrt{-x_1} \, ({x_2}-x_1) \, I^{A}_{2,1,1,1,0,0,1,0,0}\,,\nn\\
f^A_{32} &= \ep^4 ({x_2}-x_1) I^{A}_{1,1,1,0,0,1,1,0,0}\,,\nn\\
f^A_{33} &= \ep^3 \sqrt{4 {}-x_1} \, \sqrt{-x_1} \, ({x_2}-x_1) \, I^{A}_{1,2,1,0,0,1,1,0,0}\,,\nn\\
f^A_{34} &= \ep^3 \sqrt{4 {}-{x_2}} \, \sqrt{-{x_2}} \, x_1 \, I^{A}_{1,1,0,2,1,0,1,0,0}\,,\nn\\
f^A_{35} &= \ep^4 x_3 I^{A}_{1,0,1,0,1,1,1,0,0}\,,\nn\\
f^A_{36} &= \ep^4 x_1 I^{A}_{1,0,1,1,1,1,0,0,0}\,,\nn\\
f^A_{37} &= \ep^3 \sqrt{4 {}-x_1} \, \sqrt{-x_1} \, x_1 \, I^{A}_{1,0,1,2,1,1,0,0,0}\,,\nn\\
f^A_{38} &= \ep^4 ({x_2}-x_1) I^{A}_{1,1,0,1,0,1,1,0,0}\,,\nn\\
f^A_{39} &= \ep^3 \sqrt{4 {}-x_1} \, \sqrt{-x_1} \, ({x_2}-x_1) \, I^{A}_{2,1,0,1,0,1,1,0,0}\,,\nn\\
f^A_{40} &= \ep^3 \sqrt{4 {}-{x_2}} \sqrt{-{x_2}} ({x_2}-x_1) I^{A}_{1,1,0,1,0,1,2,0,0}\,,\nn\\
f^A_{41} &= \ep^2 \bigg( \ep {x_2} (x_1-{x_2}) I^{A}_{1,1,0,1,0,1,2,0,0} - \ep x_1 (x_1-{x_2}) I^{A}_{2,1,0,1,0,1,1,0,0} \nn \\
& \;\;\;\; +  (x_1-{x_2})^2 I^{A}_{2,1,0,1,0,1,2,0,0} + 2 \big( {x_2} x_1-2 {} ({x_2}+x_1) \big) I^{A}_{2,1,0,2,0,0,1,0,0} \bigg)\,,\nn\\
f^A_{42} &= \ep^4 ({x_2}-x_3) I^{A}_{0,1,0,1,1,1,1,0,0}\,,\nn\\
f^A_{43} &= \ep^3 \sqrt{4 {}-{x_2}} \sqrt{-{x_2}} ({x_2}-x_3) I^{A}_{0,1,0,1,1,1,2,0,0}\,,\nn\\
f^A_{44} &= \ep^4 ({x_2}-x_1) x_1 I^{A}_{1,1,1,1,1,0,1,0,0}\,,\nn\\
f^A_{45} &= \ep^3 \sqrt{ ({x_2}-x_1)^2+x_1^2 x_3^2 + 2 {} x_1 x_3 ({x_2}-x_1-2 x_3)} I^{A}_{0,0,1,1,2,1,1,0,0}\,,\nn\\
f^A_{46} &= \ep^2 \, \sqrt{-x_1} \, \sqrt{-x_3} \, \sqrt{4 {} ({x_2}-x_1-x_3)+x_1 x_3} \, \left( \ep I^{A}_{0,0,1,1,2,1,1,0,0} -  I^{A}_{0,0,1,1,3,1,1,0,0} \right)\,,\nn\\
f^A_{47} &= \ep^3 ({x_2}-x_1) I^{A}_{0,0,1,1,2,1,1,-1,0}\,,\nn\\
f^A_{48} &= \ep^4 ({x_2}-x_1-x_3) I^{A}_{0,1,1,0,1,1,1,0,0}\,,\nn\\
f^A_{49} &= \ep^3 \sqrt{-x_1} \, \sqrt{-x_3} \, \sqrt{x_1 x_3 + 4 {} ({x_2}-x_1-x_3)} \, I^{A}_{0,1,1,0,1,2,1,0,0}\,,\nn\\
f^A_{50} &= \ep^3 {} ({x_2}-x_1-x_3) \Big( I^{A}_{0,1,1,0,2,1,1,0,0} + I^{A}_{0,2,1,0,1,1,1,0,0} \Big)\,,\nn\\
f^A_{51} &= \ep^3 {} ({x_2}-x_1-x_3) \Big( I^{A}_{0,1,1,0,1,1,2,0,0} + I^{A}_{0,1,2,0,1,1,1,0,0} \Big)\,,\nn\\
f^A_{52} &= \ep^3 \sqrt{-x_1} \, \sqrt{- \big(  x_1 + x_1 x_3^2 + 2 {} x_3 (2 {x_2} - x_1 - 2x_3) \big)} \, I^{A}_{1,1,0,0,1,1,2,0,0}\,,\nn\\
f^A_{53} &= \ep^2 \sqrt{-x_1} \, \sqrt{-x_3} \, \sqrt{4 {} ({x_2}-x_1-x_3)+x_1 x_3} \, \Big(  I^{A}_{1,1,0,0,1,1,3,0,0} - \ep I^{A}_{1,1,0,0,1,1,2,0,0} \Big)\,,\nn\\
f^A_{54} &= \ep^3 x_1 I^{A}_{1,1,0,0,1,1,2,0,-1}\,,\nn\\
f^A_{55} &= \ep^4 \sqrt{-x_1} \, \sqrt{-x_3} \, \sqrt{4 {} ({x_2}-x_1-x_3)+x_1 x_3} \, I^{A}_{0,1,1,1,1,1,1,0,0}\,,\nn\\
f^A_{56} &= -\ep^4 \Big( (2 {x_2} - 2x_1 - x_3) I^{A}_{0,1,1,0,1,1,1,0,0} + (x_1-{x_2}) I^{A}_{0,1,1,1,1,1,1,-1,0} + {} (x_1-{x_2}) I^{A}_{0,1,1,1,1,1,1,0,0} \Big)\,,\nn\\
f^A_{57} &= \ep^2 \Big( 2 {} ({x_2}+x_1) I^{A}_{0,0,0,1,3,1,1,0,0} + \ep (x_1+x_3) I^{A}_{1,0,-1,1,1,2,1,0,0} -2 \ep ({x_2}+x_1) I^{A}_{0,0,0,1,2,1,1,0,0}\Big)\,,\nn\\
f^A_{58} &= \ep^4 (x_1+x_3) I^{A}_{1,0,0,1,1,1,1,0,0}\,,\nn\\
f^A_{59} &= \ep^3 \sqrt{4 {}-{x_2}} \, \sqrt{-{x_2}} \, \Big( x_1 I^{A}_{1,0,0,2,1,1,1,0,0} - x_3 I^{A}_{1,0,0,1,1,1,2,0,0} \Big)\,,\nn\\
f^A_{60} &= \ep^3 \sqrt{-x_1} \, \sqrt{-x_3} \, \sqrt{x_1 x_3-4 {} (-{x_2}+x_1+x_3)} \, I^{A}_{1,0,0,1,1,2,1,0,0}\,,\nn\\
f^A_{61} &= \ep^3 \big({} (x_1+x_3) - {x_2} x_3/2 \big) \Big( I^{A}_{1,0,0,1,1,1,2,0,0} + I^{A}_{1,0,0,2,1,1,1,0,0} \Big)\,,\nn\\
f^A_{62} &= \ep^4 \sqrt{-x_1} \, \sqrt{-x_3} \, \sqrt{x_1 x_3-4 {} (-{x_2}+x_1+x_3)} \, I^{A}_{1,0,1,1,1,1,1,0,0}\,,\nn\\
f^A_{63} &= \ep^4 \Big( ({x_2}+x_3) I^{A}_{1,0,0,1,1,1,1,0,0} + (x_1-{x_2}) I^{A}_{1,0,1,1,1,1,1,-1,0} + {} (x_1-{x_2}) I^{A}_{1,0,1,1,1,1,1,0,0} \Big)\,,\nn\\
f^A_{64} &= \ep^4 \sqrt{-x_1} \, \sqrt{-x_3} \, \sqrt{4 {} ({x_2}-x_1-x_3)+x_1 x_3} \, I^{A}_{1,1,1,0,1,1,1,0,0}\,,\nn\\
f^A_{65} &= \ep^4 \Big( x_1 I^{A}_{1,1,1,0,1,1,1,0,-1} + {} x_1 I^{A}_{1,1,1,0,1,1,1,0,0} - ({x_2}-x_3) I^{A}_{0,1,1,0,1,1,1,0,0} \Big)\,,\nn\\
f^A_{66} &= \ep^4 (-x_1)^{3/2} \, I^{A}_{1,1,0,1,1,1,1,0,0}\,,\nn\\
f^A_{67} &= \ep^4 (-x_1)^{3/2} \, x_1 \, I^{A}_{2,1,0,1,1,1,1,0,0}\,,\nn\\
f^A_{68} &= \ep^4 (-x_1)^{3/2} \, x_1 \, I^{A}_{1,2,0,1,1,1,1,0,0}\,,\nn\\
f^A_{69} &= \ep^4 (-x_1)^{3/2} \, I^{A}_{1,1,0,1,1,1,2,-1,0}\,,\nn\\
f^A_{70} &= \ep^4 x_1 \, \sqrt{-x_3} \, \sqrt{4 {}-x_1} \, \sqrt{4 {} ({x_2}-x_1-x_3)+x_1 x_3} \, I^{A}_{1,1,1,1,1,1,1,0,0}\,,\nn\\
f^A_{71} &= \ep^4 \sqrt{4 {}-x_1} \, \sqrt{-x_1} \, \bigg( ({x_2}-x_1) \, \Big( I^{A}_{1,1,1,1,1,1,1,-1,0} + {} I^{A}_{1,1,1,1,1,1,1,0,0} \Big) \nn \\
& \;\;\;\; - x_3 \, \frac{\sqrt{4 {} ({x_2}-x_1-x_3)+x_1 x_3}}{\sqrt{4  {x_2 - x_3} - x_1(4  - x_3)}} I^{A}_{1,1,1,0,1,1,1,0,0} + 4 \frac{x_1 - x_2}{4  - x_1} I^{A}_{1,1,0,1,1,1,1,0,0} \bigg)\,,\nn\\
f^A_{72} &= \ep^4 \sqrt{-x_1} \sqrt{4  - x_1} \bigg( x_1 \Big(  I^{A}_{1,1,1,1,1,1,1,0,0} + I^{A}_{1,1,1,1,1,1,1,0,-1} \Big) \nn \\ 
& \;\;\;\; + \frac{\sqrt{4 {} ({x_2}-x_1-x_3)+x_1 x_3}}{\sqrt{4  ( {x_2} - x_3) - x_1 (4  - x_3)}} \Big( x_3 I^{A}_{1,0,1,1,1,1,1,0,0} + (x_3 - x_2) I^{A}_{0,1,1,1,1,1,1,0,0} \Big) \bigg)\,,\nn\\
f^A_{73} &= \ep^4 \bigg( \frac{x_1}{2} \Big( (2 {}+{x_2}-2 x_1) \big( {} I^{A}_{1,1,1,1,1,1,1,0,0} + I^{A}_{1,1,1,1,1,1,1,-1,0} \big) \nn \\
& \;\;\;\;\;\;\;\;\;\; + (2 {}-x_1) I^{A}_{1,1,1,1,1,1,1,0,-1} + 2 I^{A}_{1,1,1,1,1,1,1,-1,-1} \Big) - 2 (x_1+x_3) I^{A}_{1,1,0,1,1,1,1,0,0} \nn \\
& \;\;\;\;\;\;\; + \frac{x_1 \, \sqrt{4 {} ({x_2}-x_1-x_3)+x_1 x_3}}{2 \sqrt{4  ( {x_2} - x_3) - x_1 (4  - x_3)}} \Big( ({x_2}-x_3) I^{A}_{0,1,1,1,1,1,1,0,0} \nn \\
& \;\;\;\;\;\;\;\;\;\; - x_3 \big( I^{A}_{1,0,1,1,1,1,1,0,0} + I^{A}_{1,1,1,0,1,1,1,0,0} \big) \Big) \bigg) \nn \\
& \;\;\;\; + \ep^3 \, \frac{x_1}{4} \bigg( 2 x_1 \big( I^{A}_{1,0,1,2,1,1,0,0,0} - I^{A}_{1,1,2,1,1,0,0,0,0} \big) \nn \\
& \;\;\;\;\;\;\; + (x_1-x_2) \Big( I^{A}_{1,2,1,0,0,1,1,0,0} + I^{A}_{2,1,0,1,0,1,1,0,0} - 2 I^{A}_{2,1,1,1,0,0,1,0,0} \Big) \bigg).
\end{align}

Family B is defined by the nine propagators,
\begin{gather}
 d_1^{B} = -k_1^2, \quad  d_2^{B}  = -(k_1+p_{12})^2, \quad d_3^{B} = m^2-k_2^2\,,\nn\\
 d_4^{B} = m^2-(k_2+p_{12})^2,\quad d_5^{B}  = -(k_1+p_{1})^2, \quad d_6^{B} = m^2-(k_1-k_2)^2, \\
 d_7^{B} = m^2-(k_2-p_3)^2,\quad  d_8^{B}  = m^2-(k_2+p_1)^2,\quad d_9^{B} = -(k_1-p_3)^2,\nn 
\end{gather}
with the extra restriction that $a_8$ and $a_9$ are non-positive. The family contains 50 master integrals. Below, we give the basis transformation between pre-canonical and canonical forms.
\begin{align}
f^B_{1} &= \ep^2 I^B_{0,0,0,0,0,2,2,0,0}\,,\nn\\
f^B_{2} &= \ep^2 x_1  I^B_{1,2,0,0,0,0,2,0,0}\,,\nn\\
f^B_{3} &= \ep^2 x_1  I^B_{0,1,2,0,0,2,0,0,0}\,,\nn\\
f^B_{4} &= \ep^2 \sqrt{4 {}-x_1 } \, \sqrt{-x_1 } \, \Big( I^B_{0,1,2,0,0,2,0,0,0}/2 + I^B_{0,2,2,0,0,1,0,0,0} \Big)\,,\nn\\
f^B_{5} &= \ep^2 {x_2 } I^B_{0,1,0,0,0,2,2,0,0}\,,\nn\\
f^B_{6} &= \ep^2 \sqrt{4 {}-{x_2 }} \, \sqrt{-{x_2 }} \, \Big( I^B_{0,1,0,0,0,2,2,0,0}/2 + I^B_{0,2,0,0,0,2,1,0,0} \Big)\,,\nn\\
f^B_{7} &= \ep^2 \sqrt{4 {}-x_1 } \, \sqrt{-x_1 } \, I^B_{0,0,1,2,0,2,0,0,0}\,,\nn\\
f^B_{8} &= \ep^2 \sqrt{4 {}-{x_2 }} \sqrt{-{x_2 }} I^B_{0,0,0,2,0,2,1,0,0}\,,\nn\\
f^B_{9} &= \ep^2 x_3  I^B_{0,0,0,0,1,2,2,0,0}\,,\nn\\
f^B_{10} &= \ep^2 \sqrt{4 {}-x_3 } \, \sqrt{-x_3 } \, \Big( I^B_{0,0,0,0,1,2,2,0,0}/2 + I^B_{0,0,0,0,2,2,1,0,0} \Big)\,,\nn\\
f^B_{11} &= \ep^2 \sqrt{4 {}-x_1 } \, \sqrt{-x_1 } \, x_1  I^B_{1,2,1,2,0,0,0,0,0}\,,\nn\\
f^B_{12} &= \ep^2 \sqrt{4 {}-{x_2 }} \, \sqrt{-{x_2 }} \, x_1  I^B_{1,2,0,2,0,0,1,0,0}\,,\nn\\
f^B_{13} &= \ep^3 ({x_2 }-x_1 ) I^B_{1,1,0,0,0,2,1,0,0}\,,\nn\\
f^B_{14} &= \ep^2 \frac{\sqrt{4 {} + x_1  - {x_2 }}}{\sqrt{x_1 -x_2 }} \, \Big( x_1  I^B_{1,2,0,0,0,2,1,0,-1} - x_2  I^B_{0,2,0,0,0,2,1,0,0} - \ep (x_1  - x_2 ) I^B_{1,1,0,0,0,2,1,0,0} \Big)\,,\nn\\
f^B_{15} &= \ep^3 ({x_2 }-x_1 ) I^B_{1,0,0,1,0,2,1,0,0}\,,\nn\\
f^B_{16} &= \ep^2 {} ({x_2 }-x_1 ) I^B_{1,0,0,1,0,3,1,0,0}\,,\nn\\
f^B_{17} &= \ep^2 \frac{\sqrt{4 {}-{x_2 }} \, \sqrt{-{x_2 }}}{4 ({x_2 }-2 x_1 )} \bigg( 6 \ep (x_1  - x_2 ) I^B_{1,0,0,1,0,2,1,0,0} - 4  (x_1  - x_2 ) I^B_{1,0,0,1,0,3,1,0,0} \nn \\
& \;\;\;\; + 4 \big( {} {x_2 } + x_1  (x_1 -{x_2 }) \big) I^B_{1,0,0,2,0,2,1,0,0} - 3 x_1  I^B_{0,1,2,0,0,2,0,0,0} \bigg)\,,\nn\\
f^B_{18} &= \ep^3 ({x_2 }-x_1 ) I^B_{0,1,1,0,0,2,1,0,0}\,,\nn\\
f^B_{19} &= \ep^3 x_1  I^B_{0,0,1,1,1,2,0,0,0}\,,\nn\\
f^B_{20} &= \ep^2 {} x_1  I^B_{0,0,1,1,1,3,0,0,0}\,,\nn\\
f^B_{21} &= \ep^2 \sqrt{4 {}-x_1 } \, \sqrt{-x_1 } \, \Big( I^B_{0,0,1,2,1,2,0,-1,0} - \ep I^B_{0,0,1,1,1,2,0,0,0} \Big)\,,\nn\\
f^B_{22} &= \ep^3 ({x_2 }-x_1 ) I^B_{0,0,1,1,0,2,1,0,0}\,,\nn\\
f^B_{23} &= \ep^3 x_3  I^B_{0,0,1,0,1,2,1,0,0}\,,\nn\\
f^B_{24} &= \ep^3 ({x_2 }-x_3 ) I^B_{0,0,0,1,1,2,1,0,0}\,,\nn\\
f^B_{25} &= \ep^2 {} ({x_2 }-x_3 ) I^B_{0,0,0,1,1,3,1,0,0}\,,\nn\\
f^B_{26} &= \ep^2 \frac{\sqrt{4 {}-{x_2 }}}{\sqrt{-{x_2 }} \, \big( x_3  (2 {x_2 }-x_3 ) - 2 {} ({x_2 }+x_3 ) \big)} \bigg( x_2  \big( {} {x_2 } - x_3  ({x_2 }-x_3 ) \big) I^B_{0,0,0,2,1,2,1,-1,0} \nn \\
& \;\;\;\; +  x_3 ^2 ({x_2 }-x_3 ) I^B_{0,0,0,1,1,3,1,0,0} \, + \, x_3  (4 {} {x_2 } - x_3  (4 {x_2 }-x_3 )) I^B_{0,0,0,0,1,2,2,0,0}/4 \nn \\
& \;\;\;\; - \ep \big( 2 {} {x_2 } ({x_2 }+x_3 )-x_3  (2 {x_2^2 }-3 {x_2 } x_3 +x_3 ^2) \big) I^B_{0,0,0,1,1,2,1,0,0}/2 \bigg)\,,\nn\\
f^B_{27} &= \ep^3 (1 - 2 \ep) x_1  I^B_{1,1,1,1,0,1,0,0,0}\,,\nn\\
f^B_{28} &= \ep^3 ({x_2 }-x_1 ) x_1  I^B_{1,2,1,1,0,0,1,0,0}\,,\nn\\
f^B_{29} &= \ep^4 ({x_2 }-x_1 ) I^B_{1,1,1,0,0,1,1,0,0}\,,\nn\\
f^B_{30} &= \ep^4 ({x_2 }-x_1 ) I^B_{1,1,0,1,0,1,1,0,0}\,,\nn\\
f^B_{31} &= \ep^2 {} x_1  I^B_{1,1,0,1,0,2,1,0,0} + \ep^3 (4 {}-{x_2 }) ({x_2 }+x_1 ) I^B_{1,1,0,1,0,1,2,0,0}/2 \nn \\
& \;\;\;\; - 2 \ep^4 {x_2 } I^B_{1,1,0,1,0,1,1,0,0} \,+\, \frac{\ep^2}{2(x_2 -x_1 )} \Big( \big( {x_2 } ({x_2 }-x_1 )-4 {} ({x_2 }+x_1 ) \big) I^B_{0,2,0,0,0,2,1,0,0}\nn \\
& \;\;\;\;\;\;\;\; + 2 x_1  (4 {}-{x_2 }+x_1 ) I^B_{1,2,0,0,0,2,1,0,-1} \Big) + \ep^3 (4 {}-3 {x_2 }+x_1 ) I^B_{1,1,0,0,0,2,1,0,0}\nn \\
& \;\;\;\; + \frac{\ep^2}{4(x_2 -2x_1 )} \bigg( 4 \big( 4  {x_2 }+{x_2 } ({x_2 }-x_1 ) x_1 -{} (x_2^2  + 4 {x_2 } x_1 -4 x_1 ^2) \big) I^B_{1,0,0,2,0,2,1,0,0} \nn \\
& \;\;\;\;\;\;\;\; + {} (4 {}-{x_2 }) ({x_2 }-x_1 ) I^B_{1,0,0,1,0,3,1,0,0} - 3 (4 {}-{x_2 }) x_1  I^B_{0,1,2,0,0,2,0,0,0}  \nn \\
& \;\;\;\;\;\;\;\; + 2 \ep \big( {x_2 } (5 {x_2 }-7 x_1 )-12 {} ({x_2 }-x_1 ) \big) I^B_{1,0,0,1,0,2,1,0,0} \bigg) \nn \\
& \;\;\;\; + \ep^2 ({}-{x_2 }/4) \Big( I^B_{0,1,0,0,0,2,2,0,0} - 2 I^B_{0,0,0,2,0,2,1,0,0} + 4 x_1  I^B_{1,2,0,2,0,0,1,0,0} \Big)\,,\nn\\
%
f^B_{32} &= \ep^3 \sqrt{4 {}-{x_2 }} \, \sqrt{-{x_2 }} \, ({x_2 }-x_1 ) I^B_{1,1,0,1,0,1,2,0,0}\,,\nn\\
f^B_{33} &= \ep^3 x_1  \sqrt{4 {}-x_3 } \, \sqrt{-x_3 } \, I^B_{1,1,0,0,1,2,1,0,0}\,,\nn\\
f^B_{34} &= \ep^3 x_1  \Big( I^B_{1,1,0,0,1,2,1,0,-1} + x_3  I^B_{1,1,0,0,1,2,1,0,0} \Big)\,,\nn\\
f^B_{35} &= \ep^4 (x_1 +x_3 ) I^B_{1,0,0,1,1,1,1,0,0}\,,\nn\\
f^B_{36} &= \ep^3 \sqrt{-x_1 } \, \sqrt{-x_3 } \, \sqrt{4 {} ({x_2 }-x_1 -x_3 )+x_1  x_3 } \, I^B_{1,0,0,1,1,2,1,0,0}\,,\nn\\
f^B_{37} &= \ep^2 \frac{2 {} (x_1 +x_3 ) - {x_2 } x_3 }{4 x_2  x_3  \big(2 {} ({x_2 }+x_3 ) - x_3  (2 {x_2 }-x_3 )\big)} \bigg( x_3  \big( 4 {} {x_2 } - x_3  (4 {x_2 }-x_3 ) \big) I^B_{0,0,0,0,1,2,2,0,0} \nn \\
&\;\;\;\; + 2 \ep \big( x_3  (2 x_2^2 -3 {x_2 } x_3 +x_3 ^2) -2 {} {x_2 } ({x_2 }+x_3 ) \big) I^B_{0,0,0,1,1,2,1,0,0}  \nn \\
&\;\;\;\; + 4 {} ({x_2 }-x_3 ) x_3 ^2 I^B_{0,0,0,1,1,3,1,0,0} + 4 x_2  ({} {x_2 } - x_3  ({x_2 }-x_3 )) I^B_{0,0,0,2,1,2,1,-1,0} \bigg) \nn \\
&\;\; + \ep^2 \frac{2  (x_1 +x_3 ) - x_2  x_3 }{4 x_3  ({x_2 }-2 x_1 )} \bigg( 3 x_1  I^B_{0,1,2,0,0,2,0,0,0} + 6 \ep ({x_2 }-x_1 ) I^B_{1,0,0,1,0,2,1,0,0} \nn \\
&\;\;\;\; - 4  ({x_2 }-x_1 ) I^B_{1,0,0,1,0,3,1,0,0} - 4 \big( {} {x_2 } - x_1  ({x_2 }-x_1 ) \big) I^B_{1,0,0,2,0,2,1,0,0} \bigg) \nn \\
&\;\; + \ep^3 \bigg( \big(2 {} (x_1 +x_3 ) - x_1  x_3  \big) I^B_{1,0,0,1,1,2,1,0,0}/2 \, + \,  \big({} (x_1 +x_3 )^2 - {x_2 } x_1  x_3  \big) I^B_{1,0,0,2,1,1,1,0,0}/x_3  \bigg) \nn \\
%
f^B_{38} &= \ep^3 \sqrt{ ({x_2 }-x_1 )^2+x_1 ^2 x_3 ^2 + 2 {} x_1  x_3  ({x_2 }-x_1 -2 x_3 )} \, I^B_{0,0,1,1,1,2,1,0,0}\,,\nn\\
f^B_{39} &= \ep^2 \sqrt{-x_1 } \, \sqrt{-x_3 } \, \sqrt{4 {} ({x_2 }-x_1 -x_3 )+x_1  x_3 } \, \Big(  I^B_{0,0,1,1,1,3,1,0,0} - \ep I^B_{0,0,1,1,1,2,1,0,0} \Big) \nn \\
f^B_{40} &= \ep^3 ({x_2 }-x_1 ) \Big( I^B_{0,0,1,1,1,2,1,-1,0} - {} I^B_{0,0,1,1,1,2,1,0,0} \Big)\,,\nn\\
f^B_{41} &= \ep^4 \sqrt{4 {}-x_1 } \, \sqrt{-x_1 } \, ({x_2 }-x_1 ) I^B_{1,1,1,1,0,1,1,0,0})\,,\nn\\
f^B_{42} &= \ep^4 ({x_2 }-x_1 -x_3 ) I^B_{0,1,1,0,1,1,1,0,0}\,,\nn\\
f^B_{43} &= \ep^3 \sqrt{x_1  x_3  (4 {} ({x_2 }-x_1 -x_3 )+x_1  x_3 )} \, I^B_{0,1,1,0,1,2,1,0,0}\,,\nn\\
f^B_{44} &= \ep^4 x_1  x_3  I^B_{1,1,1,0,1,1,1,0,0}\,,\nn\\
f^B_{45} &= \ep^4 x_1  ({x_2 }-x_3 ) I^B_{1,1,0,1,1,1,1,0,0}\,,\nn\\
f^B_{46} &= \ep^2 x_1  \sqrt{4 -x_2 } \sqrt{-x_2 } \Big( 2 I^B_{1,1,0,0,1,2,1,0,0} - I^B_{1,0,0,1,1,2,1,0,0} + (x_2 -x_3 ) I^B_{1,1,0,1,1,1,2,0,0} \Big)\,,\nn\\
%
f^B_{47} &= \ep^4 x_1  \sqrt{-x_1 } \, \sqrt{-x_3 } \, \sqrt{4 {} ({x_2 }-x_1 -x_3 )+x_1  x_3 } \, I^B_{1,1,1,1,1,1,1,0,0}\,,\nn\\
f^B_{48} &= \ep^4 ({x_2 }-x_1 ) x_1  I^B_{1,1,1,1,1,1,1,-1,0}\,,\nn\\
f^B_{49} &= \ep^2 \sqrt{-x_1 } \sqrt{4  - x_1 } \bigg( \ep^2 x_1 x_3  I^B_{1,1,1,1,1,1,1,0,0} + \ep^2 x_1  I^B_{1,1,1,1,1,1,1,0,-1} + \ep x_3  I^B_{1,0,0,1,1,2,1,0,0}/2 \nn \\
&\;\;\;\; - \ep (x_2 -x_3 ) I^B_{0,1,1,0,1,2,1,0,0}/2 + ({x_2 }-2 x_3 ) \Big( \ep I^B_{0,0,1,1,1,2,1,0,0} -  I^B_{0,0,1,1,1,3,1,0,0} \Big) \bigg)\,,\nn
%
\\
f^B_{50} &= 2 \ep^4 x_1  \Big( 2 I^B_{1,1,1,1,1,1,1,-1,-1} + 2 ({x_2 }-x_1 ) I^B_{1,1,1,1,1,1,1,-1,0} - x_1  I^B_{1,1,1,1,1,1,1,0,-1} \nn \\
& \;\;\;\; - x_1  x_3  I^B_{1,1,1,1,1,1,1,0,0} \Big) + \ep^2 \frac{x_2 }{x_2 -x_1 } \Big( x_2  I^B_{0,1,0,0,0,2,2,0,0} - x_1  I^B_{0,1,2,0,0,2,0,0,0} \Big) \nn \\
& \;\; -2 \ep^3 {x_2 } \Big( I^B_{0,0,1,1,0,2,1,0,0} -2 I^B_{0,0,1,1,1,2,1,-1,0} + I^B_{0,1,1,0,0,2,1,0,0} \Big) \nn \\
& \;\; 2 \ep^2 {} x_1  ({x_2 }-2 x_3 ) I^B_{0,0,1,1,1,3,1,0,0} - 2 \ep^3 \big( 2 {} {x_2 }+x_1  ({x_2 }-2 x_3 ) \big) I^B_{0,0,1,1,1,2,1,0,0} \nn \\
& \;\; + \ep^3 x_1  \Big( ({x_2 }-x_3 ) I^B_{0,1,1,0,1,2,1,0,0} - x_3  I^B_{1,0,0,1,1,2,1,0,0} - 4 x_3  I^B_{1,2,1,1,0,0,1,0,0} \Big)  \nn \\
& \;\; - 4 \ep^4 x_1  \Big( I^B_{0,1,1,0,1,1,1,0,0} + I^B_{1,0,0,1,1,1,1,0,0} + I^B_{1,1,1,1,0,1,0,0,0} \Big) \nn \\
& \;\; + 4 \ep^4 {x_2 } I^B_{1,1,0,1,0,1,1,0,0} - 2 \ep^4 ({x_2 }-x_1 ) x_1  I^B_{1,1,1,1,0,1,1,0,0}. 
%
\end{align}

Family C is defined by the nine propagators,
\begin{gather}
 d_1^{C} = -k_1^2, \quad d_2^{C} = -(k_1+p_{12})^2, \quad d_3^{C} = m^2-(k_2+p_{12})^2,\nn  \\
 d_4^{C} = -(k_1+p_{1})^2, \quad d_5^{C} = m^2-(k_1-k_2)^2, \quad d_6^{C} = m^2-(k_2-p_3)^2, \\
 d_7^{C} = -(k_1-p_3)^2, \quad d_8^{C} = m^2-k_2^2, \quad d_9^{C} = m^2-(k_2+p_1)^2\,,\nn
\end{gather}
with the extra restriction that $a_8$ and $a_9$ are non-positive. The family contains 45 master integrals. Below, we give the basis transformation between pre-canonical and canonical forms.
\begin{align}
f^C_{1} &= \ep^2 I^C_{0,0,0,0,2,2,0,0,0}\,,\nn\\
f^C_{2} &= \ep^2 x_3  I^C_{0,0,0,1,0,2,2,0,0}\,,\nn\\
f^C_{3} &= \ep^2 x_3  I^C_{0,0,0,1,2,2,0,0,0}\,,\nn\\
f^C_{4} &= \ep^2 \sqrt{4 {}-x_3 } \, \sqrt{-x_3 } \, \Big( I^C_{0,0,0,2,1,2,0,0,0} + I^C_{0,0,0,1,2,2,0,0,0}/2 \Big)\,,\nn\\
f^C_{5} &= \ep^2 {x_2 } I^C_{0,0,2,0,2,0,1,0,0}\,,\nn\\
f^C_{6} &= \ep^2 \sqrt{4 {}-{x_2 }} \, \sqrt{-{x_2 }} \, \Big( I^C_{0,0,1,0,2,0,2,0,0} + I^C_{0,0,2,0,2,0,1,0,0}/2 \Big)\,,\nn\\
f^C_{7} &= \ep^2 \sqrt{4 {}-{x_2 }} \, \sqrt{-{x_2 }} \, I^C_{0,0,1,0,2,2,0,0,0}\,,\nn\\
f^C_{8} &= \ep^2 {x_2 } I^C_{0,1,0,0,0,2,2,0,0}\,,\nn\\
f^C_{9} &= \ep^2 x_1  I^C_{1,0,2,0,2,0,0,0,0}\,,\nn\\
f^C_{10} &= \ep^2 \sqrt{4 {}-x_1 } \, \sqrt{-x_1 } \, \Big( I^C_{2,0,1,0,2,0,0,0,0} + I^C_{1,0,2,0,2,0,0,0,0}/2 \Big)\,,\nn\\
f^C_{11} &= \ep^2 x_1  I^C_{1,2,0,0,0,2,0,0,0}\,,\nn\\
f^C_{12} &= \ep^2 x_3  \sqrt{4 {}-{x_2 }} \, \sqrt{-{x_2 }} \, I^C_{0,0,1,1,0,2,2,0,0}\,,\nn\\
f^C_{13} &= \ep^3 ({x_2 }-x_3 ) I^C_{0,0,1,1,2,0,1,0,0}\,,\nn\\
f^C_{14} &= \ep^2 \frac{\sqrt{4 {}-{x_2 }+x_3 }}{\sqrt{x_3 -{x_2 }}} \Big( x_3  I^C_{0,-1,1,1,2,0,2,0,0} - x_2  I^C_{0,0,1,0,2,0,2,0,0} - \ep (x_3  - x_2 ) I^C_{0,0,1,1,2,0,1,0,0} \Big)\,,\nn\\
f^C_{15} &= \ep^3 ({x_2 }-x_3 ) I^C_{0,0,1,1,2,1,0,0,0}\,,\nn\\
f^C_{16} &= \ep^2 {} ({x_2 }-x_3 ) I^C_{0,0,1,1,3,1,0,0,0}\,,\nn\\
f^C_{17} &= \ep^2 \frac{\sqrt{4 {}-{x_2 }} \sqrt{-{x_2 }}}{4 ({x_2 }-2 x_3 )} \bigg( 4  ({x_2 }-x_3 ) I^C_{0,0,1,1,3,1,0,0,0} - 6 \ep ({x_2 }-x_3 ) I^C_{0,0,1,1,2,1,0,0,0} \nn \\
& \;\;\;\; + 4 \big( {} {x_2 } - x_3  ({x_2 }-x_3 ) \big) I^C_{0,0,1,1,2,2,0,0,0} - 3 x_3  I^C_{0,0,0,1,2,2,0,0,0} \bigg)\,,\nn\\
%
f^C_{18} &= \ep^2 {x_2 } \sqrt{4 {}-{x_2 }} \, \sqrt{-{x_2 }} \, I^C_{0,1,1,0,0,2,2,0,0}\,,\nn\\
f^C_{19} &= \ep^3 ({x_2 }-x_1 ) I^C_{1,0,1,0,2,1,0,0,0}\,,\nn\\
f^C_{20} &= \ep^2 {} ({x_2 }-x_1 ) I^C_{1,0,1,0,3,1,0,0,0}\,,\nn\\
f^C_{21} &= \ep^2 \frac{\sqrt{4 {}-{x_2 }} \sqrt{-{x_2 }}}{4 ({x_2 }-2 x_1 )} \bigg( 4  ({x_2 }-x_1 ) I^C_{1,0,1,0,3,1,0,0,0} - 6 \ep ({x_2 }-x_1 ) I^C_{1,0,1,0,2,1,0,0,0} \nn \\
& \;\;\;\; + 4 \big( {} {x_2 } - x_1  ({x_2 }-x_1 ) \big) I^C_{1,0,2,0,2,1,0,0,0} - 3 x_1  I^C_{1,0,2,0,2,0,0,0,0} \bigg)\,,\nn
%
\\
f^C_{22} &= \ep^3 ({x_2 }-x_1 ) I^C_{1,1,0,0,2,1,0,0,0}\,,\nn\\
f^C_{23} &= \ep^2 \frac{\sqrt{4 {}-{x_2 }+x_1 }}{\sqrt{x_1 -{x_2 }}} \Big( x_1  I^C_{1,2,0,0,2,1,-1,0,0} - {x_2 } I^C_{0,2,0,0,2,1,0,0,0} - \ep (x_1  - x_2 ) I^C_{1,1,0,0,2,1,0,0,0} \Big),   \nn \\
%
f^C_{24} &= \ep^2 x_1  \sqrt{4 {}-{x_2 }} \, \sqrt{-{x_2 }} \, I^C_{1,2,1,0,0,2,0,0,0}\,,\nn\\
f^C_{25} &= \ep^4 ({x_2 }-x_3 ) I^C_{0,0,1,1,1,1,1,0,0}\,,\nn\\
f^C_{26} &= \ep^2 {} x_3  I^C_{0,0,1,1,2,1,1,0,0} + \ep^3 (4 {}-{x_2 }) ({x_2 }+x_3 ) I^C_{0,0,2,1,1,1,1,0,0}/2 \nn \\
& \;\; - 2 \ep^4 {x_2 } I^C_{0,0,1,1,1,1,1,0,0} + \frac{\ep^2}{2 (x_2 -x_3 )} \Big( \big( {x_2 } ({x_2 }-x_3 )-4 {} ({x_2 }+x_3 ) \big) I^C_{0,0,1,0,2,0,2,0,0} \nn \\
& \;\;\;\; + 2 x_3  (4 {}-{x_2 }+x_3 ) I^C_{0,-1,1,1,2,0,2,0,0} \Big) + \ep^3 (4 {}-3 {x_2 }+x_3 ) I^C_{0,0,1,1,2,0,1,0,0} \nn \\
& \;\; + \frac{\ep^2}{4 ({x_2 }-2 x_3 )} \bigg( 2 \ep \big( {x_2 } (5 {x_2 }-7 x_3 )-12 {} ({x_2 }-x_3 ) \big) I^C_{0,0,1,1,2,1,0,0,0} \nn \\
& \;\;\;\; + 4 \big( 4  {x_2 }+{x_2 } ({x_2 }-x_3 ) x_3 -{} (x_2^2 +4 {x_2 } x_3 -4 x_3 ^2) \big) I^C_{0,0,1,1,2,2,0,0,0} \nn \\
& \;\;\;\; + 4 {} (4 {}-{x_2 }) ({x_2 }-x_3 ) I^C_{0,0,1,1,3,1,0,0,0} - 3 (4 {}-{x_2 }) x_3  I^C_{0,0,0,1,2,2,0,0,0} \bigg) \nn \\
& \;\; + \ep^2 (4  -x_2 ) \Big( x_3  I^C_{0,0,1,1,0,2,2,0,0} - I^C_{0,0,1,0,2,2,0,0,0}/2 + I^C_{0,0,2,0,2,0,1,0,0}/4 \Big)\,,\nn \\
%
f^C_{27} &= \ep^3 \sqrt{4 {}-{x_2 }} \, \sqrt{-{x_2 }} \, ({x_2 }-x_3 ) I^C_{0,0,2,1,1,1,1,0,0}\,,\nn\\
f^C_{28} &= (1-2 \ep) \ep^3 {x_2 } I^C_{0,1,1,0,1,1,1,0,0}\,,\nn\\
f^C_{29} &= \ep^3 \sqrt{4 {}-x_1 } \, \sqrt{-x_1 } \, x_3  I^C_{1,0,1,1,2,0,1,0,0}\,,\nn\\
f^C_{30} &= \ep^3 x_3  \Big( I^C_{1,-1,1,1,2,0,1,0,0} + x_1  I^C_{1,0,1,1,2,0,1,0,0} \Big)\,,\nn\\
f^C_{31} &= \ep^4 (x_1 +x_3 ) I^C_{1,0,1,1,1,1,0,0,0}\,,\nn\\
f^C_{32} &= \ep^3 \sqrt{-x_1 } \, \sqrt{-x_3 } \, \sqrt{4 {} ({x_2 }-x_1 -x_3 )+x_1  x_3 } \, I^C_{1,0,1,1,2,1,0,0,0}\,,\nn\\
f^C_{33} &= \ep^2 \frac{2 {} (x_1 +x_3 ) - {x_2 } x_3 }{4 {x_2 } x_3  \big( 2 {} ({x_2 }+x_3 ) - x_3  (2 {x_2 }-x_3 ) \big)} \bigg( x_3  \big( 4 {} {x_2 } - x_3  (4 {x_2 }-x_3 ) \big) I^C_{0,0,0,1,2,2,0,0,0} \nn \\
& \;\;\;\; - 2 \ep \big( 2 {} {x_2 } ({x_2 }+x_3 ) - x_3  (2 x_2^2  -3 {x_2 } x_3 +x_3 ^2) \big) I^C_{0,0,1,1,2,1,0,0,0} \nn \\
& \;\;\;\; + 4 {} ({x_2 }-x_3 ) x_3 ^2 I^C_{0,0,1,1,3,1,0,0,0} + 4 x_2  \big( {} {x_2 } - x_3  ({x_2 }-x_3 ) \big) I^C_{0,0,2,1,2,1,0,0,-1} \bigg) \nn \\
& \;\; + \ep^2 \frac{2 {} (x_1 +x_3 ) - {x_2 } x_3 }{4 (x_2  - 2x_1 ) x_3 } \bigg( 3 x_1  I^C_{1,0,2,0,2,0,0,0,0} - 4 {} ({x_2 }-x_1 ) I^C_{1,0,1,0,3,1,0,0,0} \nn \\
& \;\;\;\; + 6 \ep ({x_2 }-x_1 ) I^C_{1,0,1,0,2,1,0,0,0} - 4 \big( {} {x_2 }+x_1  (-{x_2 }+x_1 ) \big) I^C_{1,0,2,0,2,1,0,0,0} \bigg) \nn \\
& \;\; + \ep^3 \Big( \big(2 {} (x_1 +x_3 ) - x_1  x_3  \big) I^C_{1,0,1,1,2,1,0,0,0}/2  +  \big( {} (x_1 +x_3 )^2 - {x_2 } x_1  x_3  \big) I^C_{1,0,2,1,1,1,0,0,0})/x_3  \Big)\,,\nn
%
\\
f^C_{34} &= \ep^3 x_1  x_3  I^C_{1,1,0,1,0,2,1,0,0}\,,\nn\\
f^C_{35} &= \ep^3 x_1  \sqrt{4 {}-x_3 } \, \sqrt{-x_3 } \, I^C_{1,1,0,1,2,1,0,0,0}\,,\nn\\
f^C_{36} &= \ep^3 x_1  \Big( I^C_{1,1,0,1,2,1,-1,0,0} + x_3  I^C_{1,1,0,1,2,1,0,0,0} \Big)\,,\nn\\
f^C_{37} &= \ep^4 ({x_2 }-x_1 ) I^C_{1,1,1,0,1,1,0,0,0}\,,\nn\\
f^C_{38} &= \ep^2 {} x_1  I^C_{1,1,1,0,2,1,0,0,0} + \ep^3 (4 {}-{x_2 }) ({x_2 }+x_1 ) I^C_{1,1,1,0,1,2,0,0,0}/2  \nn \\
& \;\; - 2 \ep^4 {x_2 } I^C_{1,1,1,0,1,1,0,0,0} + \frac{\ep^2}{2 ({x_2 }-x_1 )} \bigg( \big( {x_2 } ({x_2 }-x_1 )-4 {} ({x_2 }+x_1 ) \big) I^C_{0,2,0,0,2,1,0,0,0} \nn \\
& \;\;\;\; + 2 x_1  (4 {}-{x_2 }+x_1 ) I^C_{1,2,0,0,2,1,-1,0,0} \bigg) + \ep^3 (4 {}-3 {x_2 }+x_1 ) I^C_{1,1,0,0,2,1,0,0,0} \nn \\
& \;\; + \frac{\ep^2}{4 ({x_2 }-2 x_1 )} \bigg( 2 \ep \big( {x_2 } (5 {x_2 }-7 x_1 )-12 {} ({x_2 }-x_1 ) \big) I^C_{1,0,1,0,2,1,0,0,0} \nn \\
& \;\;\;\; + 4 \big( 4  {x_2 }+{x_2 } ({x_2 }-x_1 ) x_1 -{} (x_2^2  + 4 {x_2 } x_1 -4 x_1 ^2) \big) I^C_{1,0,2,0,2,1,0,0,0} \nn \\
& \;\;\;\; 4 {} (4 {}-{x_2 }) ({x_2 }-x_1 ) I^C_{1,0,1,0,3,1,0,0,0} - 3 (4 {} - {x_2 }) x_1  I^C_{1,0,2,0,2,0,0,0,0} \bigg) \nn \\
& \;\; + \ep^2 (4  -x_2 ) \Big( I^C_{0,1,0,0,2,2,0,0,0}/4 - I^C_{0,0,2,0,2,1,0,0,0}/2 + x_1  I^C_{1,2,2,0,0,1,0,0,0} \Big)\,,\nn\\
%
%
f^C_{39} &= \ep^3 \sqrt{4 {}-{x_2 }} \, \sqrt{-{x_2 }} \, ({x_2 }-x_1 ) I^C_{1,1,1,0,1,2,0,0,0}\,,\nn\\
f^C_{40} &= \ep^4 ({x_2 }-x_1 ) x_3  I^C_{1,0,1,1,1,1,1,0,0}\,,\nn\\
f^C_{41} &= \ep^3 \sqrt{4 {}-{x_2 }} \, \sqrt{-{x_2 }} \, x_3  \Big( 2 I^C_{1,0,1,1,2,0,1,0,0} - I^C_{1,0,1,1,2,1,0,0,0} + ({x_2 }-x_1 ) I^C_{1,0,2,1,1,1,1,0,0} \Big)\,,\nn \\
f^C_{42} &= \ep^3 \sqrt{4 {}-{x_2 }} \, \sqrt{-{x_2 }} \, x_1  x_3  I^C_{1,1,1,1,0,2,1,0,0}\,,\nn\\
f^C_{43} &= \ep^4 x_1  ({x_2 }-x_3 ) I^C_{1,1,1,1,1,1,0,0,0}\,,\nn\\
f^C_{44} &= \ep^3 \sqrt{4 {}-{x_2 }} \, \sqrt{-{x_2 }} \, x_1  \Big( 2 I^C_{1,1,0,1,2,1,0,0,0} - I^C_{1,0,1,1,2,1,0,0,0} + ({x_2 }-x_3 ) I^C_{1,1,1,1,1,2,0,0,0} \Big)\,,\nn\\
f^C_{45} &= \ep^4 x_2  \Big( x_3  I^C_{1,0,1,1,1,1,1,0,0} + x_1  I^C_{1,1,1,1,1,1,0,0,0} + x_1  x_3  I^C_{1,1,1,1,1,1,1,0,0} \Big). 
\end{align}

Family D is defined by the nine propagators,
\begin{gather}
 d_1^{D} = m^2-k_1^2, \quad d_2^{D} = m^2-(k_1+p_{12})^2, \quad d_3^{D} = m^2-k_2^2\,,\nn \\
 d_4^{D} = m^2-(k_2+p_{12})^2, \quad d_5^{D} = m^2-(k_1+p_{1})^2, \quad d_6^{D} = -(k_1-k_2)^2, \\
 d_7^{D} = m^2-(k_2-p_3)^2, \quad d_8^{D} = m^2-(k_2+p_1)^2, \quad d_9^{D} = m^2-(k_1-p_3)^2,\nn 
\end{gather}
with the extra restriction that $a_1$, $a_5$, and $a_6$ are non-positive. The family contains 17 master integrals. Below, we give the basis transformation between pre-canonical and canonical forms.
\begin{align}
f^D_{1} &= \ep^2 I^D_{0,0,0,0,0,0,2,0,2}\,,\nn\\
f^D_{2} &= -\ep^2 \sqrt{4 {}-x_3 } \sqrt{-x_3 } I^D_{0,0,0,0,0,0,1,2,2}\,,\nn\\
f^D_{3} &= -\ep^2 \sqrt{4 {}-x_1 } \sqrt{-x_1 } I^D_{0,0,2,1,0,0,0,0,2}\,,\nn\\
f^D_{4} &= -\ep^2 \sqrt{4 {}-{x_2 }} \sqrt{-{x_2 }} I^D_{0,1,2,0,0,0,0,0,2}\,,\nn\\
f^D_{5} &= \ep^3 (x_3 -{x_2 }) I^D_{0,0,0,1,0,0,1,1,2}\,,\nn\\
f^D_{6} &= \ep^3 x_3  I^D_{0,0,1,0,0,0,1,1,2}\,,\nn\\
f^D_{7} &= \ep^3 x_1  I^D_{0,0,1,1,0,0,0,1,2}\,,\nn\\
f^D_{8} &= \ep^3 (x_1 -{x_2 }) I^D_{0,0,1,1,0,0,1,0,2}\,,\nn\\
f^D_{9} &= \ep^2 \sqrt{4 {}-{x_2 }} \, \sqrt{-{x_2 }} \, \sqrt{4 {}-x_3 } \, \sqrt{-x_3 } \, I^D_{0,1,0,0,0,0,1,2,2}\,,\nn\\
f^D_{10} &= -\ep^2 {x_2 } (4 {} - {x_2 }) I^D_{0,1,0,1,0,0,2,0,2}\,,\nn\\
f^D_{11} &= \ep^2 \sqrt{4 {}-{x_2 }} \, \sqrt{-{x_2 }} \, \sqrt{4 {}-x_1 } \, \sqrt{-x_1 } \, I^D_{0,1,1,2,0,0,0,0,2}\,,\nn\\
f^D_{12} &= -\ep^3 \sqrt{-x_1 } \, \sqrt{-x_3 } \, \sqrt{4 {} ({x_2 }-x_1 -x_3 )+x_1  x_3 } \, I^D_{0,0,1,1,0,0,1,1,2})\,,\nn\\
f^D_{13} &= \ep^3 \sqrt{4 {}-{x_2 }} \, \sqrt{-{x_2 }} \, ({x_2 }-x_3 ) I^D_{0,1,0,1,0,0,1,1,2}\,,\nn\\
f^D_{14} &= \ep^3 \sqrt{4 {}-{x_2 }} \, \sqrt{-{x_2 }} \, x_3  I^D_{0,1,1,0,0,0,1,1,2}\,,\nn\\
f^D_{15} &= \ep^3 \sqrt{4 {}-{x_2 }} \, \sqrt{-{x_2 }} \, x_1  I^D_{0,1,1,1,0,0,0,1,2}\,,\nn\\
f^D_{16} &= \ep^3 \sqrt{4 {}-{x_2 }} \, \sqrt{-{x_2 }} \, ({x_2 }-x_1 ) I^D_{0,1,1,1,0,0,1,0,2}\,,\nn\\
f^D_{17} &= \ep^3 \sqrt{4 {}-{x_2 }} \, \sqrt{-{x_2 }} \, \sqrt{-x_1 } \, \sqrt{-x_3 } \, \sqrt{4 {} ({x_2 }-x_1 -x_3 )+x_1  x_3 } \, I^D_{0,1,1,1,0,0,1,1,2}. \hspace{1.25cm}
\end{align}

\section{Pre-canonical master integrals}
\label{sec:appendix:precanonical}

In this appendix we draw the 125 master integrals in the pre-canonical form and we 
link them to the corresponding integral(s) in the canonical basis.
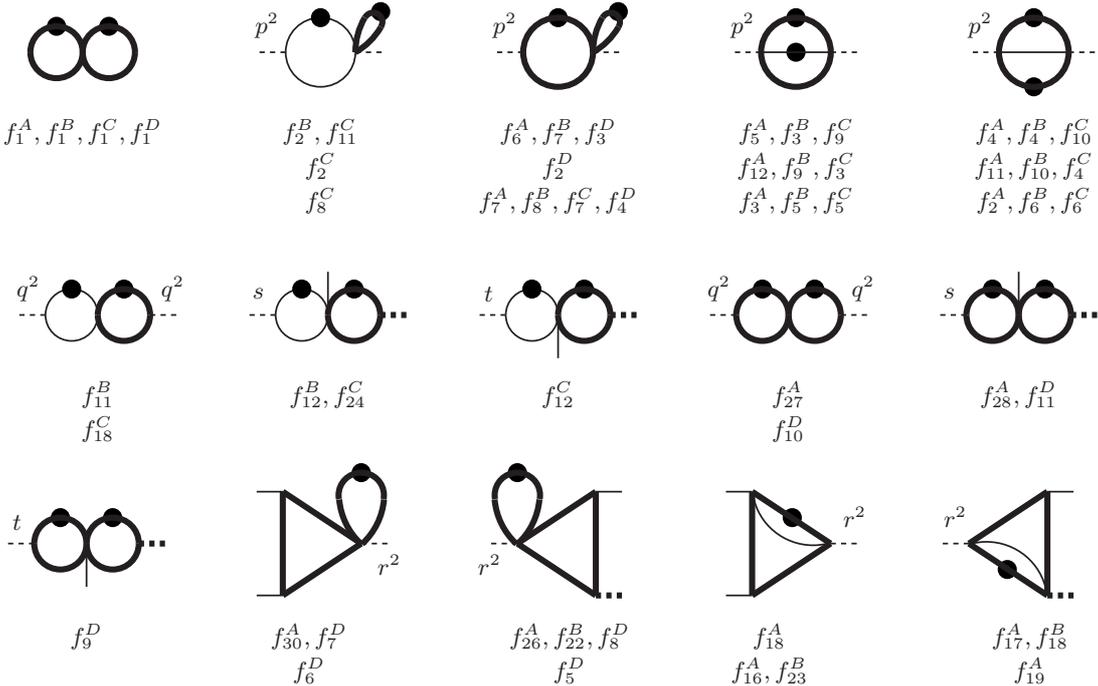
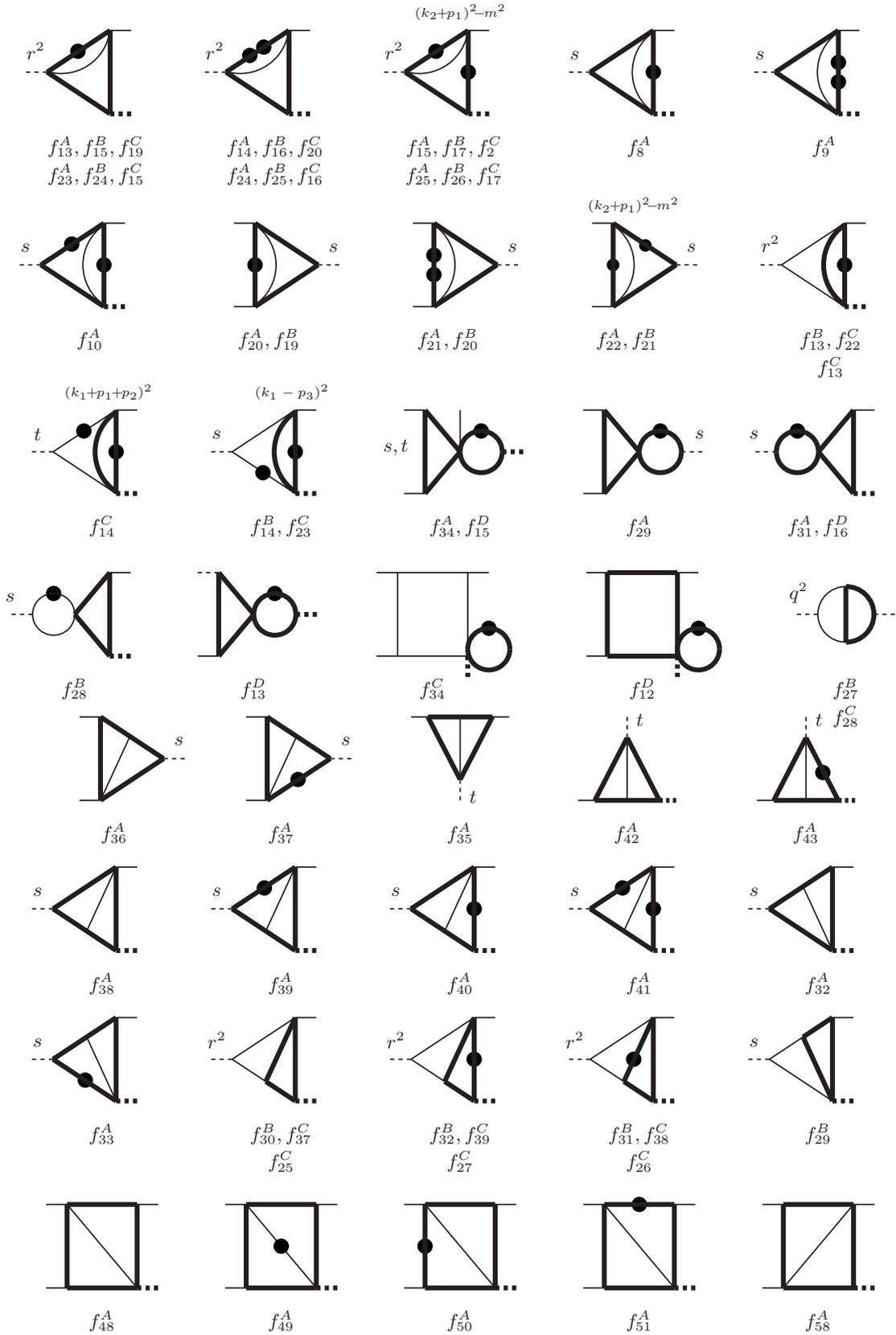
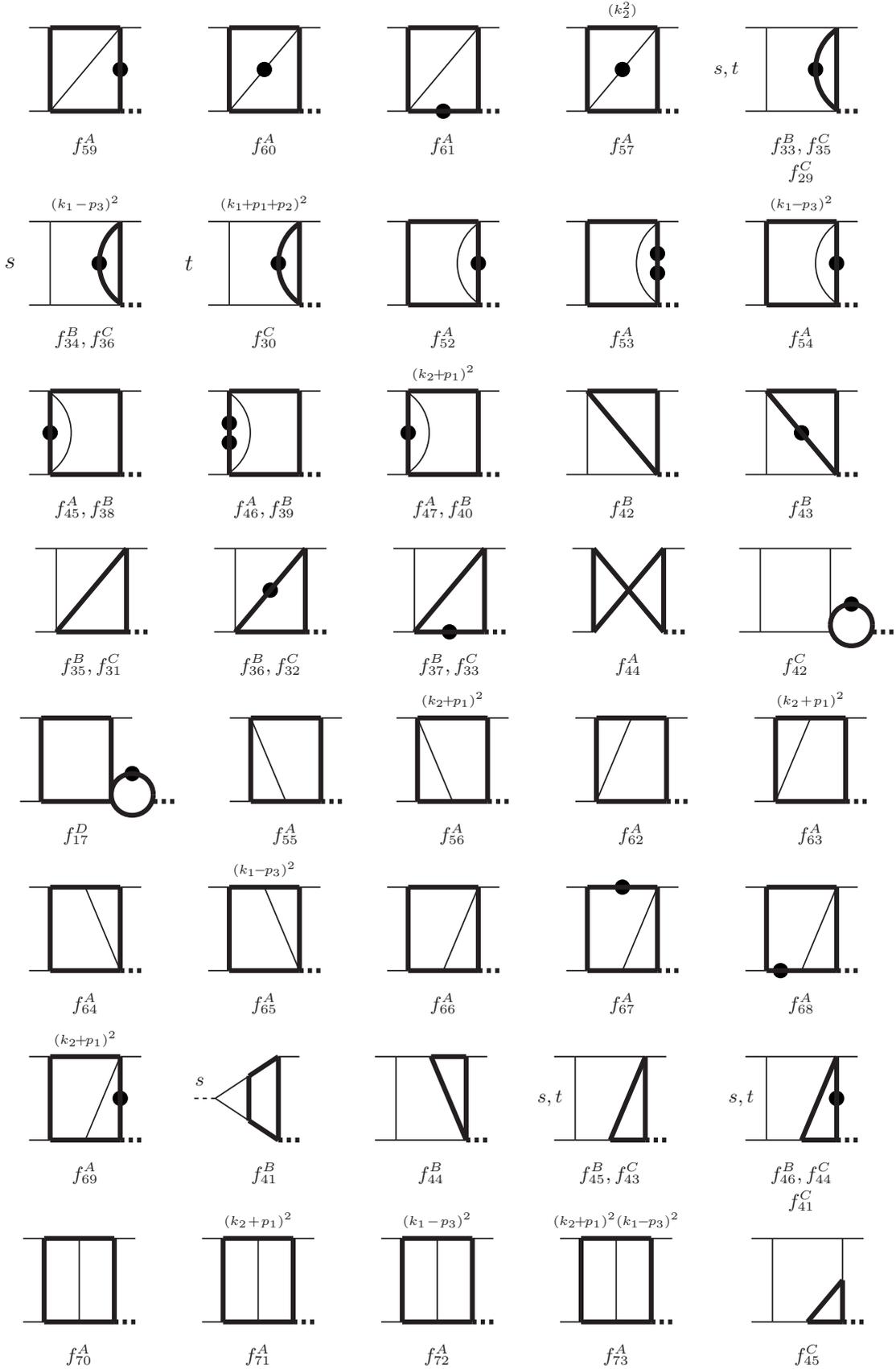
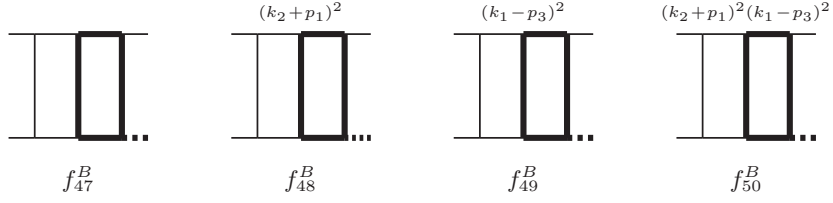
\bc
\[ \vcenter{
\hbox{
}
}
\]
\ec
%

\vspace*{15mm}

\section{Alphabet}
\label{sec:appendix:alphabet}

In this appendix we list the alphabet for the four integral families defined in section \ref{sec:integral families}. We introduce the following shorthands for the set of 13 square roots,
\begin{align}
 &R_1(x_1)=\sqrt{-x_1}\,,\,
 R_1(x_3)=\sqrt{-x_3}\,,\,
 R_1(x_2)=\sqrt{-x_2}\,,\nn\\ 
& R_2(x_1)=\sqrt{4-x_1}\,,\,
 R_2(x_3)=\sqrt{4-x_3}\,, \,
 R_2(x_2)=\sqrt{4-x_2}\,,\nn\\ 
 &R_3(x_1)=\sqrt{x_2-x_1}\,, \,
 R_3(x_3)=\sqrt{x_2-x_3}\,,\nn\\ 
 &R_4(x_1)=\sqrt{x_2-x_1-4}\,, \,
 R_4(x_3)=\sqrt{x_2-x_3-4}\,,\nn\\ 
 & R_5(x)=\sqrt{4 x_2+x_1 x_3-4 (x_1+x_3)}\,,\nn\\ 
 & R_6(x)=\sqrt{2 x_3 (-2 x_2+x_1+2 x_3)-x_1 x_3^2-x_1}\,,\nn\\
 &R_7(x)=\sqrt{2 x_1 x_3 (x_2-x_1)+(x_2-x_1)^2+(x_1-4) x_1 x_3^2}\,. \hspace{0cm}
\end{align}
They appear in the alphabet in the following 8 linearly independent combinations,
\begin{align}
&R_1(x_1) R_2(x_1)\,,\,R_1(x_2) R_2(x_2)\,,\nn\\
&R_1(x_3) R_2(x_3)\,,\,R_3(x_1)R_4(x_1)\,,\nn\\
&R_3(x_3)R_4(x_3)\,,\,R_1(x_1)R_1(x_3)R_5(x)\,,\nn\\
&R_1(x_1)R_6(x)\,,\,R_7(x).
\end{align}

Referring to the matrix $\tilde{A}$ defined in (\ref{eq:def atilde}) the alphabets of the four families can be written in terms of the following linearly independent 49 letters,

\begin{align}
&\log (x_3),\;\log (x_1),\; \log (x_2)\,,\nn\\ 
& \log (x_1-4), \;\log (x_3-4),\; \log (x_2-4)\,,\nn\\ 
& \log (x_1+x_3),\;\log (x_3-x_2), \; \log (x_1-x_2)\,,\nn\\   
& \log (-x_2+x_1+x_3),\;\log (-x_2+x_3+4),\;\log (-x_2+x_1+4)\,,\nn\\ 
&\log (4 x_2-4 x_1+x_1 x_3-4 x_3),\;\log \left(x_1^2-x_2 x_3 x_1+2 x_3 x_1+x_3^2\right)\,,\nn\\
& \log \left(x_3^2-x_2 x_3+x_2\right),\;\log \left(x_1^2-x_2x_1+x_2\right),\;\log \left(x_2^2-x_1 x_2+x_1\right)\,,\nn\\
&\log \left(x_2-x_1+x_1 x_3+R_7(x)\right)\,,\nn\\ 
& \log \left(x_2^2-x_1 x_2+x_1 x_3 x_2-2 x_1 x_3^2+R_7(x) x_2\right)\,,\nn\\ 
& \log \left(-x_3 x_1^2+x_1^2-x_2 x_1+3 x_3 x_1-R_7(x) x_1+x_1-x_2+R_7(x)\right)\,,\nn\\
&\log \left(x_3 x_1-x_1-2 x_2 x_3+R_1(x_1) R_6(x)\right)\,,\nn\\
&\log \left(x_2 x_1 x_1-2 x_1-2 x_3+R_1(x_2) R_2(x_2)\right)\,,\nn\\
& \log \left(x_3-R_1(x_3) R_2(x_3)\right),\; \log \left(x_1-R_1(x_1) R_2(x_1)\right)\,,\nn\\ 
& \log \left(-x_3 x_1^2+x_1^2-2 x_2 x_1+4 x_3 x_1+R_2(x_1) R_6(x) x_1\right)\,,\nn\\
& \log  \left(x_2-R_1(x_2) R_2(x_2)\right),\;  \log \left(x_2-x_3+R_3(x_3) R_4(x_3)\right)\,,\nn\\ 
& \log \left(x_2-x_1+R_3(x_1) R_4(x_1)\right),\;  \log \left(x_2-2 x_3+R_1(x_2) R_2(x_2)\right),\nn \nn\\ 
& \log \left(x_2-2 x_1+R_1(x_2) R_2(x_2)\right),\;  \log \left(x_3 x_1-x_1+R_1(x_1) R_6(x)\right)\,,\nn\\ 
& \log \left(-x_3 x_1-x_1+R_1(x_1) R_6(x)\right),\;  \log \left(-x_2 x_1+2 x_1+x_2 R_1(x_1) R_2(x_1)\right)\,,\nn\\
& \log \left(x_1 x_3+R_1(x_1) R_1(x_3) R_5(x)\right)\,,\nn\\ 
&  \log \left(x_3 x_1-2 x_1-2 x_3+R_1(x_1) R_1(x_3) R_5(x)\right)\,,\nn\\ 
&  \log \left(x_3 x_1^2-x_1^2+x_2 x_1-4 x_3 x_1+R_1(x_1) R_2(x_1) R_7(x)\right)\,,\nn\\ 
&  \log \left(-x_2^2+x_1 x_2-x_1 x_3 x_2+2 x_3 x_2+2 x_1 x_3+R_1(x_2) R_2(x_2) R_7(x)\right)\,,\nn\\ 
&  \log \left(-x_3^2 x_1^2+3 x_3 x_1^2+4 x_3^2 x_1-4 x_2 x_3 x_1+R_1(x_3) R_5(x) R_6(x) x_1\right)\,,\nn\\
&  \log \left(x_3 R_1(x_2) R_2(x_2)+x_2 R_1(x_3) R_2(x_3)\right)\,,\nn\\ 
&\log \left(x_1 R_1(x_2) R_2(x_2)+x_2 R_1(x_1) R_2(x_1)\right)\,,\nn\\ 
& \log \left(x_1   R_1(x_3) R_2(x_3)-R_1(x_1) R_1(x_3) R_5(x)\right)\,,\nn\\ 
& \log \left(x_3 R_1(x_1) R_2(x_1)-R_1(x_1) R_1(x_3) R_5(x)\right)\,,\nn\\ 
& \log \left(-x_2 R_1(x_1) R_2(x_1)+x_3 R_1(x_1) R_2(x_1)+x_1 R_3(x_3) R_4(x_3)\right)\,,\nn\\ 
& \log \left(-x_2 R_1(x_2) R_2(x_2)+x_3  R_1(x_2) R_2(x_2)+x_2 R_3(x_3) R_4(x_3)\right)\,,\nn\\ 
&\log \left(-x_2 R_1(x_3) R_2(x_3)+x_1 R_1(x_3) R_2(x_3)+x_3 R_3(x_1) R_4(x_1)\right)\,,\nn\\ 
& \log \left(-x_2 R_1(x_2) R_2(x_2)+x_1 R_1(x_2) R_2(x_2)+x_2 R_3(x_1) R_4(x_1)\right)\,,\nn\\ 
& \log \left(-x_3^2 x_1^2+3 x_3  x_1^2+4 x_3^2 x_1-3 x_2 x_3 x_1+R_1(x_1) R_1(x_3) R_5(x) R_7(x)\right)\,,\nn\\
& \log \left(x_2 R_1(x_1) R_1(x_3) R_5(x)-x_1 x_3 R_1(x_2) R_2(x_2)\right)\,,\nn\\ 
& \log \left(-x_2 x_3+x_1 x_3+R_1(x_2) R_2(x_2) x_3-R_1(x_1) R_1(x_3)  R_5(x)\right).
  \end{align}

\section{Weight-two functions}
\label{sec:appendix:li2}

In section \ref{sec:solution canonical} we described how to express the non-elliptic master integrals in terms of a minimal set of logarithms and dilogarithms up to weight two, while the weight three and four components are expressed as one-fold integrals over linear combinations of weight-one and weight-two functions with algebraic coefficients. 

In this appendix we list the basis choice we made for the set of linearly independent dilogarithms required to express the master integrals of each family at weight two. They are chosen to be single-valued in the Euclidean region $x_3<x_2<x_1<0$.

Family A,

\begin{align}
 & \text{Li}_{2}\left(\frac{x_{1}}{x_{1}-4}\right),\nn\\
 & \text{Li}_{2}\left(\frac{x_{2}}{x_{2}-4}\right),\nn\\
 & \text{Li}_{2}\left(\frac{x_{3}}{x_{3}-4}\right)\,,\nn\\
 & \text{Li}_{2}\left(\frac{R_{1}\left(x_{3}\right)-R_{2}\left(x_{3}\right)}{R_{1}\left(x_{3}\right)}\right),\nn\\
 & \text{Li}_{2}\left(\frac{R_{1}\left(x_{2}\right)-R_{2}\left(x_{2}\right)}{R_{1}\left(x_{2}\right)}\right),\nn\\
 & \text{Li}_{2}\left(\frac{R_{1}\left(x_{1}\right)-R_{2}\left(x_{1}\right)}{R_{1}\left(x_{1}\right)}\right),\nn\\
 & \text{Li}_{2}\left(\frac{\left(R_{1}\left(x_{3}\right)-R_{2}\left(x_{3}\right)\right){}^{2}}{\left(R_{1}\left(x_{3}\right)+R_{2}\left(x_{3}\right)\right){}^{2}}\right),\nn\\
 & \text{Li}_{2}\left(\frac{\left(R_{1}\left(x_{2}\right)-R_{2}\left(x_{2}\right)\right){}^{2}}{\left(R_{1}\left(x_{2}\right)+R_{2}\left(x_{2}\right)\right){}^{2}}\right),\nn\\
 & \text{Li}_{2}\left(\frac{\left(R_{1}\left(x_{1}\right)-R_{2}\left(x_{1}\right)\right){}^{2}}{\left(R_{1}\left(x_{1}\right)+R_{2}\left(x_{1}\right)\right){}^{2}}\right),\nn\\
 & \text{Li}_{2}\left(\frac{R_{1}\left(x_{1}\right)R_{1}\left(x_{3}\right)-R_{5}(x)}{R_{1}\left(x_{3}\right)\left(R_{1}\left(x_{1}\right)-R_{2}\left(x_{1}\right)\right)}\right),\nn\\
 & \text{Li}_{2}\left(\frac{R_{1}\left(x_{3}\right)\left(R_{1}\left(x_{2}\right)-R_{2}\left(x_{2}\right)\right)}{R_{1}\left(x_{2}\right)\left(R_{1}\left(x_{3}\right)+R_{2}\left(x_{3}\right)\right)}\right),\nn\\
 & \text{Li}_{2}\left(\frac{R_{1}\left(x_{3}\right)\left(R_{1}\left(x_{2}\right)+R_{2}\left(x_{2}\right)\right)}{R_{1}\left(x_{2}\right)\left(R_{1}\left(x_{3}\right)-R_{2}\left(x_{3}\right)\right)}\right),\nn\\
 & \text{Li}_{2}\left(\frac{R_{1}\left(x_{1}\right)\left(R_{1}\left(x_{2}\right)+R_{2}\left(x_{2}\right)\right)}{R_{1}\left(x_{2}\right)\left(R_{1}\left(x_{1}\right)-R_{2}\left(x_{1}\right)\right)}\right),\nn\\
 & \text{Li}_{2}\left(\frac{R_{1}\left(x_{1}\right)\left(R_{1}\left(x_{2}\right)-R_{2}\left(x_{2}\right)\right)}{R_{1}\left(x_{2}\right)\left(R_{1}\left(x_{1}\right)+R_{2}\left(x_{1}\right)\right)}\right),\nn\\
 & \text{Li}_{2}\left(\frac{R_{1}\left(x_{1}\right)\left(R_{1}\left(x_{3}\right)+R_{2}\left(x_{3}\right)\right)}{R_{1}\left(x_{1}\right)R_{1}\left(x_{3}\right)-R_{5}(x)}\right),\nn\\
 & \text{Li}_{2}\left(\frac{R_{1}\left(x_{3}\right)\left(R_{1}\left(x_{1}\right)+R_{2}\left(x_{1}\right)\right)}{R_{1}\left(x_{1}\right)R_{1}\left(x_{3}\right)-R_{5}(x)}\right),\nn\\
 & \text{Li}_{2}\left(-\frac{R_{1}\left(x_{1}\right)\left(R_{1}\left(x_{3}\right)-R_{2}\left(x_{3}\right)\right)}{R_{1}\left(x_{1}\right)R_{1}\left(x_{3}\right)-R_{5}(x)}\right),\nn\\
 & \text{Li}_{2}\left(\frac{R_{1}\left(x_{1}\right){}^{2}\left(R_{1}\left(x_{3}\right)-R_{2}\left(x_{3}\right)\right){}^{2}}{\left(R_{1}\left(x_{1}\right)R_{1}\left(x_{3}\right)-R_{5}(x)\right){}^{2}}\right),\nn\\
 & \text{Li}_{2}\left(\frac{R_{1}\left(x_{1}\right)\left(R_{1}\left(x_{2}\right)-R_{2}\left(x_{2}\right)\right)}{R_{1}\left(x_{2}\right)R_{2}\left(x_{1}\right)-R_{1}\left(x_{1}\right)R_{2}\left(x_{2}\right)}\right),\nn\\
 & \text{Li}_{2}\left(\frac{R_{1}\left(x_{2}\right)R_{2}\left(x_{3}\right)-R_{1}\left(x_{3}\right)R_{2}\left(x_{2}\right)}{R_{1}\left(x_{3}\right)\left(R_{1}\left(x_{2}\right)-R_{2}\left(x_{2}\right)\right)}\right),\nn\\
 & \text{Li}_{2}\left(-\frac{R_{1}\left(x_{2}\right)R_{2}\left(x_{3}\right)-R_{1}\left(x_{3}\right)R_{2}\left(x_{2}\right)}{R_{1}\left(x_{3}\right)\left(R_{1}\left(x_{2}\right)+R_{2}\left(x_{2}\right)\right)}\right),\nn\\
 & \text{Li}_{2}\left(-\frac{R_{1}\left(x_{1}\right)\left(R_{1}\left(x_{2}\right)+R_{2}\left(x_{2}\right)\right)}{R_{1}\left(x_{2}\right)R_{2}\left(x_{1}\right)-R_{1}\left(x_{1}\right)R_{2}\left(x_{2}\right)}\right),\nn\\
 & \text{Li}_{2}\left(\frac{R_{1}\left(x_{1}\right)R_{1}\left(x_{3}\right)\left(R_{1}\left(x_{2}\right)+R_{2}\left(x_{2}\right)\right)}{R_{1}\left(x_{2}\right)\left(R_{1}\left(x_{1}\right)R_{1}\left(x_{3}\right)-R_{5}(x)\right)}\right),\nn\\
 & \text{Li}_{2}\left(-\frac{R_{1}\left(x_{1}\right)R_{1}\left(x_{3}\right)\left(R_{1}\left(x_{2}\right)-R_{2}\left(x_{2}\right)\right)}{R_{1}\left(x_{2}\right)\left(R_{1}\left(x_{1}\right)R_{1}\left(x_{3}\right)-R_{5}(x)\right)}\right),\nn\\
 & \text{Li}_{2}\left(\frac{R_{1}\left(x_{1}\right){}^{2}R_{1}\left(x_{3}\right){}^{2}\left(R_{1}\left(x_{2}\right)-R_{2}\left(x_{2}\right)\right){}^{2}}{R_{1}\left(x_{2}\right){}^{2}\left(R_{1}\left(x_{1}\right)R_{1}\left(x_{3}\right)-R_{5}(x)\right){}^{2}}\right).\hspace{2.5cm}
\end{align}

Family B,

\begin{align}
 & \text{Li}_{2}\left(\frac{x_{1}}{x_{1}-4}\right),\nn\\
 & \text{Li}_{2}\left(\frac{x_{2}}{x_{2}-4}\right),\nn\\
 & \text{Li}_{2}\left(\frac{x_{3}}{x_{3}-4}\right)\,,\nn\\
 & \text{Li}_{2}\left(\frac{R_{1}\left(x_{3}\right)-R_{2}\left(x_{3}\right)}{R_{1}\left(x_{3}\right)}\right),\nn\\
 & \text{Li}_{2}\left(\frac{R_{1}\left(x_{2}\right)-R_{2}\left(x_{2}\right)}{R_{1}\left(x_{2}\right)}\right),\nn\\
 & \text{Li}_{2}\left(\frac{R_{1}\left(x_{1}\right)-R_{2}\left(x_{1}\right)}{R_{1}\left(x_{1}\right)}\right),\nn\\
 & \text{Li}_{2}\left(\frac{\left(R_{1}\left(x_{3}\right)-R_{2}\left(x_{3}\right)\right){}^{2}}{\left(R_{1}\left(x_{3}\right)+R_{2}\left(x_{3}\right)\right){}^{2}}\right),\nn\\
 & \text{Li}_{2}\left(\frac{\left(R_{1}\left(x_{2}\right)-R_{2}\left(x_{2}\right)\right){}^{2}}{\left(R_{1}\left(x_{2}\right)+R_{2}\left(x_{2}\right)\right){}^{2}}\right),\nn\\
 & \text{Li}_{2}\left(\frac{\left(R_{1}\left(x_{1}\right)-R_{2}\left(x_{1}\right)\right){}^{2}}{\left(R_{1}\left(x_{1}\right)+R_{2}\left(x_{1}\right)\right){}^{2}}\right),\nn\\
 & \text{Li}_{2}\left(\frac{R_{1}\left(x_{3}\right)\left(R_{1}\left(x_{2}\right)-R_{2}\left(x_{2}\right)\right)}{R_{1}\left(x_{2}\right)\left(R_{1}\left(x_{3}\right)+R_{2}\left(x_{3}\right)\right)}\right),\nn\\
 & \text{Li}_{2}\left(\frac{R_{1}\left(x_{3}\right)\left(R_{1}\left(x_{2}\right)+R_{2}\left(x_{2}\right)\right)}{R_{1}\left(x_{2}\right)\left(R_{1}\left(x_{3}\right)-R_{2}\left(x_{3}\right)\right)}\right),\nn\\
 & \text{Li}_{2}\left(\frac{R_{1}\left(x_{1}\right)\left(R_{1}\left(x_{2}\right)+R_{2}\left(x_{2}\right)\right)}{R_{1}\left(x_{2}\right)\left(R_{1}\left(x_{1}\right)-R_{2}\left(x_{1}\right)\right)}\right),\nn\\
 & \text{Li}_{2}\left(\frac{R_{1}\left(x_{1}\right)\left(R_{1}\left(x_{2}\right)-R_{2}\left(x_{2}\right)\right)}{R_{1}\left(x_{2}\right)\left(R_{1}\left(x_{1}\right)+R_{2}\left(x_{1}\right)\right)}\right),\nn\\
 & \text{Li}_{2}\left(\frac{x_{2}+R_{1}\left(x_{2}\right)R_{2}\left(x_{2}\right)-2}{-x_{1}+x_{2}+R_{3}\left(x\right)R_{4}\left(x\right)-2}\right),\nn\\
 & \text{Li}_{2}\left(\frac{R_{1}\left(x_{1}\right)\left(R_{1}\left(x_{2}\right)-R_{2}\left(x_{2}\right)\right)}{R_{1}\left(x_{2}\right)R_{2}\left(x_{1}\right)-R_{1}\left(x_{1}\right)R_{2}\left(x_{2}\right)}\right),\nn\\
 & \text{Li}_{2}\left(\frac{R_{1}\left(x_{2}\right)R_{2}\left(x_{3}\right)-R_{1}\left(x_{3}\right)R_{2}\left(x_{2}\right)}{R_{1}\left(x_{3}\right)\left(R_{1}\left(x_{2}\right)-R_{2}\left(x_{2}\right)\right)}\right),\nn\\
 & \text{Li}_{2}\left(-\frac{R_{1}\left(x_{1}\right)\left(R_{1}\left(x_{2}\right)+R_{2}\left(x_{2}\right)\right)}{R_{1}\left(x_{2}\right)R_{2}\left(x_{1}\right)-R_{1}\left(x_{1}\right)R_{2}\left(x_{2}\right)}\right),\nn\\
 & \text{Li}_{2}\left(-\frac{R_{1}\left(x_{2}\right)R_{2}\left(x_{3}\right)-R_{1}\left(x_{3}\right)R_{2}\left(x_{2}\right)}{R_{1}\left(x_{3}\right)\left(R_{1}\left(x_{2}\right)+R_{2}\left(x_{2}\right)\right)}\right),\nn\\
 & \text{Li}_{2}\left(\frac{R_{1}\left(x_{1}\right)R_{1}\left(x_{3}\right)\left(x_{3}-R_{1}\left(x_{3}\right)R_{2}\left(x_{3}\right)\right)}{x_{3}\left(R_{1}\left(x_{1}\right)R_{1}\left(x_{3}\right)-R_{5}(x)\right)}\right),\nn\\
 & \text{Li}_{2}\left(\frac{R_{1}\left(x_{1}\right)R_{1}\left(x_{3}\right)\left(x_{2}-R_{1}\left(x_{2}\right)R_{2}\left(x_{2}\right)\right)}{x_{2}\left(R_{1}\left(x_{1}\right)R_{1}\left(x_{3}\right)-R_{5}(x)\right)}\right),\nn\\
 & \text{Li}_{2}\left(\frac{R_{1}\left(x_{1}\right)R_{1}\left(x_{3}\right)\left(x_{1}-R_{1}\left(x_{1}\right)R_{2}\left(x_{1}\right)\right)}{x_{1}\left(R_{1}\left(x_{1}\right)R_{1}\left(x_{3}\right)-R_{5}(x)\right)}\right),\nn\\
 & \text{Li}_{2}\left(\frac{R_{1}\left(x_{1}\right)R_{1}\left(x_{3}\right)\left(R_{1}\left(x_{2}\right)-R_{2}\left(x_{2}\right)\right)}{R_{1}\left(x_{2}\right)R_{5}(x)-R_{1}\left(x_{1}\right)R_{1}\left(x_{3}\right)R_{2}\left(x_{2}\right)}\right),\nn\\
 & \text{Li}_{2}\left(-\frac{R_{1}\left(x_{1}\right)R_{1}\left(x_{3}\right)\left(x_{1}+R_{1}\left(x_{1}\right)R_{2}\left(x_{1}\right)\right)}{x_{1}\left(R_{1}\left(x_{1}\right)R_{1}\left(x_{3}\right)-R_{5}(x)\right)}\right),\nn\\
 & \text{Li}_{2}\left(-\frac{R_{1}\left(x_{1}\right)R_{1}\left(x_{3}\right)\left(x_{3}+R_{1}\left(x_{3}\right)R_{2}\left(x_{3}\right)\right)}{x_{3}\left(R_{1}\left(x_{1}\right)R_{1}\left(x_{3}\right)-R_{5}(x)\right)}\right),\nn\\
 & \text{Li}_{2}\left(-\frac{x_{1}\left(R_{1}\left(x_{1}\right)R_{1}\left(x_{3}\right)-R_{5}(x)\right){}^{2}}{R_{1}\left(x_{1}\right){}^{2}R_{1}\left(x_{3}\right){}^{2}\left(R_{1}\left(x_{1}\right)-R_{2}\left(x_{1}\right)\right){}^{2}}\right),\nn\\
 & \text{Li}_{2}\left(\frac{2R_{1}\left(x_{1}\right){}^{2}R_{1}\left(x_{3}\right){}^{2}\left(x_{3}+R_{1}\left(x_{3}\right)R_{2}\left(x_{3}\right)-2\right)}{x_{3}\left(R_{1}\left(x_{1}\right)R_{1}\left(x_{3}\right)-R_{5}(x)\right){}^{2}}\right),\nn\\
 & \text{Li}_{2}\left(\frac{x_{1}x_{2}\left(R_{1}\left(x_{2}\right)+R_{2}\left(x_{2}\right)\right)}{-R_{3}\left(x\right)R_{4}\left(x\right)R_{1}\left(x_{2}\right){}^{3}-x_{2}^{2}R_{2}\left(x_{2}\right)+x_{1}x_{2}R_{2}\left(x_{2}\right)}\right),\nn\\
\hspace{0.5cm} & \text{Li}_{2}\left(-\frac{\left(x_{1}-x_{2}\right)x_{2}\left(R_{1}\left(x_{2}\right)-R_{2}\left(x_{2}\right)\right)}{-R_{3}\left(x\right)R_{4}\left(x\right)R_{1}\left(x_{2}\right){}^{3}-x_{2}^{2}R_{2}\left(x_{2}\right)+x_{1}x_{2}R_{2}\left(x_{2}\right)}\right).\hspace{.7cm}
\end{align}

Family C,

\begin{align}
& \Li_2 \left( \frac{ 1-x_2 }{ x_1} \right),\nn\\
& \Li_2 \left( \frac{ 1-x_2 }{ x_3} \right),\nn\\
& \Li_2 \left( \frac{ x_1 }{ x_1-x_2+x_3 } \right),\nn\\
& \Li_2 \left( \frac{ x_1 x_3 }{ x_2 (x_1-x_2+x_3) } \right),\nn\\
& \Li_2 \left( \frac{ (R_1(x_2)+R_2(x_2))^2 }{ (R_1(x_3)+R_2(x_3))^2} \right) ,\nn\\
& \Li_2 \left( \frac{ (R_1(x_1)+R_2(x_1))^2 }{ (R_1(x_2)+R_2(x_2))^2} \right),\nn\\
& \Li_2 \left( \frac{ -4 }{ (R_3(x_1)+R_4(x_1))^2} \right),\nn\\
& \Li_2 \left( \frac{ 16 }{ (R_1(x_3)+R_2(x_3))^4} \right),\nn\\
& \Li_2 \left( \frac{ -4 }{ (R_1(x_3)+R_2(x_3))^2} \right),\nn\\
& \Li_2 \left( \frac{ 16 }{ (R_1(x_2)+R_2(x_2))^4} \right),\nn\\
& \Li_2 \left( \frac{ -4 }{ (R_1(x_2)+R_2(x_2))^2} \right),\nn\\
& \Li_2 \left( \frac{ 16 }{ (R_1(x_1)+R_2(x_1))^4} \right),\nn\\
& \Li_2 \left( \frac{ -4 }{ (R_1(x_1)+R_2(x_1))^2} \right),\nn\\
& \Li_2 \left( \frac{ -4 }{ (R_3(x_3)+R_4(x_3))^2} \right),\nn\\
& \Li_2 \left( \frac{ -4 (x_1-x_2+x_3) }{ (R_1(x_1) R_1(x_3)+R_5(x))^2} \right),\nn\\
& \Li_2 \left( \frac{ R_1(x_1) (R_1(x_3)+R_2(x_3)) }{ R_1(x_1) R_1(x_3)+R_5(x) } \right),\nn\\
& \Li_2 \left( \frac{ R_1(x_2) (R_1(x_2)+R_2(x_2)) }{ R_1(x_3) (R_1(x_3)+R_2(x_3)) } \right),\nn\\
& \Li_2 \left( \frac{ R_1(x_1) R_1(x_3) (R_1(x_2)+R_2(x_2)) }{ R_1(x_2) (R_1(x_1) R_1(x_3)+R_5(x)) } \right) ,\nn\\
& \Li_2 \left( \frac{ - x_1 (R_1(x_2)+R_2(x_2))^2 }{ (R_2(x_2) R_3(x_1)+R_1(x_2) R_4(x_1))^2 } \right),\nn\\
& \Li_2 \left( \frac{ -(x_1-x_2+x_3) (R_1(x_1)+R_2(x_1))^2 }{ R_1(x_1) R_1(x_3)+R_5(x))^2 } \right),\nn\\
& \Li_2 \left( \frac{ 16 }{ (R_1(x_1)+R_2(x_1))^2 (R_1(x_2)+R_2(x_2))^2} \right),\nn\\
& \Li_2 \left( \frac{ 16 }{ (R_1(x_2)+R_2(x_2))^2 (R_1(x_3)+R_2(x_3))^2 } \right),\nn\\
& \Li_2 \left( \frac{ -x_1 (R_1(x_2)+R_2(x_2)) }{ R_3(x_1) (R_2(x_2) R_3(x_1)+R_1(x_2) R_4(x_1))} \right),\nn\\
& \Li_2 \left( \frac{ -4 R_1(x_1) }{ (R_1(x_3)+R_2(x_3)) (R_1(x_1) R_1(x_3)+R_5(x)) } \right),\nn\\
& \Li_2 \left( \frac{ -4 R_1(x_2) }{ R_1(x_3) (R_1(x_2)+R_2(x_2)) (R_1(x_3)+R_2(x_3)) } \right),\nn\\
& \Li_2 \left( \frac{ -16 (x_1-x_2+x_3) }{ (R_1(x_1)+R_2(x_1))^2 (R_1(x_1) R_1(x_3)+R_5(x))^2 } \right),\nn\\
& \Li_2 \left( \frac{ -4 R_1(x_1) R_1(x_3) }{ R_1(x_2) (R_1(x_2)+R_2(x_2)) (R_1(x_1) R_1(x_3)+R_5(x)) } \right),\nn\\
& \Li_2 \left( \frac{ 4 R_3(x_3) }{ (R_1(x_2)+R_2(x_2)) (R_2(x_2) R_3(x_3)+R_1(x_2) R_4(x_3)) } \right),\nn\\
& \Li_2 \left( \frac{ -16 x_1 }{ (R_1(x_2)+R_2(x_2))^2 (R_2(x_2) R_3(x_1)+R_1(x_2) R_4(x_1))^2} \right),\nn\\
& \Li_2 \left( \frac{ -4 x_1 }{ (R_1(x_2)+R_2(x_2)) R_3(x_1) (R_2(x_2) R_3(x_1)+R_1(x_2) R_4(x_1)) } \right),\nn\\
\hspace{1.3cm}& \Li_2 \left( \frac{ -4 x_3 }{ (R_1(x_2)+R_2(x_2)) R_3(x_3) (R_2(x_2) R_3(x_3)+R_1(x_2) R_4(x_3)) } \right).
\end{align}

Family D,

\begin{align}
 & \text{Li}_{2}\left(\frac{x_{2}-4}{x_{3}-4}\right),\nn\\
 & \text{Li}_{2}\left(\frac{x_{1}-4}{x_{2}-4}\right),\nn\\
 & \text{Li}_{2}\left(\frac{R_{1}\left(x_{3}\right)}{R_{2}\left(x_{3}\right)}\right),\nn\\
 & \text{Li}_{2}\left(-\frac{R_{1}\left(x_{3}\right)}{R_{2}\left(x_{3}\right)}\right),\nn\\
 & \text{Li}_{2}\left(\frac{R_{5}(x)}{R_{1}\left(x_{3}\right)R_{2}\left(x_{1}\right)}\right),\nn\\
 & \text{Li}_{2}\left(-\frac{R_{5}(x)}{R_{1}\left(x_{3}\right)R_{2}\left(x_{1}\right)}\right),\nn\\
 & \text{Li}_{2}\left(\frac{R_{2}\left(x_{2}\right)}{R_{1}\left(x_{2}\right)+R_{2}\left(x_{2}\right)}\right),\nn\\
  & \text{Li}_{2}\left(\frac{R_{2}\left(x_{1}\right)}{R_{1}\left(x_{1}\right)+R_{2}\left(x_{1}\right)}\right),\nn\\
 & \text{Li}_{2}\left(-\frac{4\left(x_{2}-4\right)}{\left(x_{1}-4\right)\left(x_{3}-4\right)}\right),\nn\\
 & \text{Li}_{2}\left(\frac{16}{\left(R_{1}\left(x_{1}\right)+R_{2}\left(x_{1}\right)\right){}^{4}}\right),\nn\\
 & \text{Li}_{2}\left(\frac{16}{\left(R_{1}\left(x_{2}\right)+R_{2}\left(x_{2}\right)\right){}^{4}}\right),\nn\\
 & \text{Li}_{2}\left(\frac{16}{\left(R_{1}\left(x_{3}\right)+R_{2}\left(x_{3}\right)\right){}^{4}}\right),\nn\\
 & \text{Li}_{2}\left(\frac{R_{1}\left(x_{1}\right)R_{1}\left(x_{3}\right)R_{2}\left(x_{2}\right)}{R_{1}\left(x_{2}\right)R_{5}(x)}\right),\nn\\
 & \text{Li}_{2}\left(\frac{R_{1}\left(x_{1}\right)R_{1}\left(x_{3}\right)+R_{5}(x)}{R_{1}\left(x_{3}\right)R_{2}\left(x_{1}\right)+R_{5}(x)}\right),\nn\\
 & \text{Li}_{2}\left(\frac{R_{1}\left(x_{1}\right)R_{1}\left(x_{3}\right)+R_{5}(x)}{R_{1}\left(x_{1}\right)R_{2}\left(x_{3}\right)+R_{5}(x)}\right),\nn\\
 & \text{Li}_{2}\left(\frac{R_{5}(x)}{R_{1}\left(x_{1}\right)R_{1}\left(x_{3}\right)+R_{5}(x)}\right),\nn\\
 & \text{Li}_{2}\left(\frac{R_{5}(x)}{R_{1}\left(x_{1}\right)R_{2}\left(x_{3}\right)+R_{5}(x)}\right),\nn\\
 & \text{Li}_{2}\left(-\frac{R_{1}\left(x_{2}\right)R_{5}(x)}{R_{1}\left(x_{1}\right)R_{1}\left(x_{3}\right)R_{2}\left(x_{2}\right)}\right),\nn\\
 & \text{Li}_{2}\left(\frac{4}{R_{2}\left(x_{1}\right)\left(R_{1}\left(x_{1}\right)+R_{2}\left(x_{1}\right)\right)}\right),\nn\\
 & \text{Li}_{2}\left(\frac{4}{R_{2}\left(x_{2}\right)\left(R_{1}\left(x_{2}\right)+R_{2}\left(x_{2}\right)\right)}\right),\nn\\
 & \text{Li}_{2}\left(\frac{R_{1}\left(x_{1}\right)R_{1}\left(x_{3}\right)+R_{5}(x)}{R_{1}\left(x_{3}\right)\left(R_{1}\left(x_{1}\right)+R_{2}\left(x_{1}\right)\right)}\right),\nn\\
 & \text{Li}_{2}\left(\frac{R_{2}\left(x_{3}\right)\left(R_{1}\left(x_{1}\right)R_{1}\left(x_{3}\right)+R_{5}(x)\right)}{\left(R_{1}\left(x_{3}\right)+R_{2}\left(x_{3}\right)\right)R_{5}(x)}\right),\nn\\
 & \text{Li}_{2}\left(\frac{R_{1}\left(x_{1}\right)R_{1}\left(x_{3}\right)\left(R_{1}\left(x_{2}\right)+R_{2}\left(x_{2}\right)\right)}{R_{1}\left(x_{2}\right)\left(R_{1}\left(x_{1}\right)R_{1}\left(x_{3}\right)+R_{5}(x)\right)}\right),\nn\\
 & \text{Li}_{2}\left(\frac{R_{1}\left(x_{2}\right)\left(R_{1}\left(x_{1}\right)R_{1}\left(x_{3}\right)+R_{5}(x)\right)}{R_{1}\left(x_{1}\right)R_{1}\left(x_{3}\right)R_{2}\left(x_{2}\right)+R_{1}\left(x_{2}\right)R_{5}(x)}\right).\hspace{2.5cm}
\end{align}

\section{One-fold integral representations}
\label{sec:appendix:integral representation}

We consider a system of differential equations for a set of integrals $f(x,\epsilon)$ in canonical form~\cite{Henn:2013pwa} defined by a matrix $\tilde{A}(x)$,
\begin{equation}
df^{(i+1)}(x)= \epsilon d \tilde{A} (x)f^{(i)}(x)\,.
\end{equation}
If some boundary values $f^{(i+1)}(0)$ and a parametrization of the integration path are provided, the equations can be readily integrated. The integration path goes from the boundary point to $x$. If the boundary point is $x=0$ a convenient parametrization is $x(\alpha)=x\, \alpha$ with $\alpha \in [0,1]$. The solution reads
\begin{equation}
\label{sol}
f^{(i+1)}(x)=\int_{0}^{1}d\alpha\,(\partial_{\alpha} \tilde{A}(\alpha))f^{(i)}(\alpha)+f^{(i+1)}(0)\,.
\end{equation}
Performing an integration by parts we can reduce the weight of the
functions involved,
\be
\begin{split}
\label{solpart}
f^{(i+1)}(x)&=\tilde{A}(1)\int_{0}^{1} d\alpha\,(\partial_{\alpha}\tilde{A}(\alpha))f^{(i-1)}(\alpha)-\int_{0}^{1}d\alpha\, \tilde{A}(\alpha)(\partial_{\alpha} \tilde{A}(\alpha))f^{(i-1)}(\alpha)\\
&+ \int_{0}^{1}d\alpha\,(\partial_{\alpha}\tilde{A}(\alpha)) f^{(i)}(0)+f^{(i+1)}(0)\,.
\end{split}
\ee
If the weight-two functions are known analytically, weight-three functions can be computed
numerically using eq.~(\ref{sol}), while weight-four functions are computed via eq.~(\ref{solpart}).

If the boundary value $f(0,\epsilon)$ is singular, the integral above cannot be computed numerically since the integrand has non-integrable singularities in $\alpha=0$. However in our case all the divergent integrals are factorisable into products of one-loop integrals, which are known analytically~\cite{Gehrmann:1999as,Ellis:2007qk}. In such cases we need to define the integrals that, via $\tilde{A}(x)$, depend on the singular ones.

Assume that integral $f_{k}(x,\epsilon)$ has a singular boundary condition $f_{k}(0,\epsilon)$, and that it is known analytically to all orders of $\epsilon$. Consider an integral $f_n(x,\epsilon)$, with ${n\neq k}$, with a regular boundary condition $f_n(0,\epsilon)$. Using eq.~(\ref{sol}) we can write it as,
\begin{equation}
f_{n}^{(i+1)}(x)=\sum_{m \neq k}\int_{0}^{1}d\alpha\,(\partial_{\alpha}\tilde{A}_{nm}(\alpha))f^{(i)}_{m}(\alpha)+\int_{0}^{1}d\alpha\,(\partial_{\alpha}\tilde{A}_{n k}(\alpha))f_{k}^{(i)}(\alpha) +f_{n}^{(i+1)}(0)\,,
\end{equation}
where we made explicit the dependence on the singular integral $f_{k}^{(i)}(x)$. 
Since by assumption $f_k^{(i)}(x)$ is known analytically, we can directly compute the second integral on the right hand side. Note that the fact that $f_{n}(0,\epsilon)$ is regular ensures that the second integral is convergent even if $f_k^{(i)}(0)$ is singular. Finally we can perform an integration by parts and reduce the other integrals to the form of (\ref{solpart}).

\section{Maximal cut of the elliptic sectors} 
\label{sec:appendix:cut}

We show that the maximal cut~\cite{Cachazo:2008vp, Kosower:2011ty} of $I^{A}_{1,1,0,1,1,1,1,0,0}$ provides useful information about the class of functions needed to represent the result. We cut the six visible propagators. We parametrize the two loop momenta using the spinor-helicity formalism~\cite{Dixon:1996wi} (see~\citep{Drummond:2008vq, Hodges:2010kq} for a different formalism), 
\be 
\begin{split}
k_1^\mu & = z_1 p_1^\mu+z_2 p_2^\mu +z_3\frac{\langle 1^- | \gamma^\mu | 2^-\rangle}{2 \langle 13 \rangle [32]}+z_4 \frac{\langle 2^-| \gamma^\mu |1^-\rangle}{2 \langle 23 \rangle [31]},\\
k_2^\mu & = z_5 p_1^\mu+z_6 p_2^\mu +z_7\frac{\langle 1^- | \gamma^\mu | 2^-\rangle}{2 \langle 13 \rangle [32]}+z_8 \frac{\langle 2^-| \gamma^\mu |1^-\rangle}{2 \langle 23 \rangle [31]}\,.
\end{split}
\ee
We get the following two-fold integral result for the maximal cut,
\begin{equation}
\bar{I}=\frac{s_{13} s_{23}}{s_{12}^2}\int dx_6 dx_8 \frac{1}{\sqrt{F_1\, F_2}}\,,
\label{eq:hexacut}
\end{equation}
where the two factors under the square root are,
\be 
\begin{split}
F_1 &= m^2 s_{13} s_{23}-s_{12} z_8 \left(\left(s_{12}+s_{23}\right) z_6-s_{13}+z_8\right),\\
F_2  &= m^2 s_{13} s_{23} \left(2 z_6+1\right){}^2+4 m^2 \left(s_{12}+s_{13}\right) z_6 z_8-s_{12} z_8 \left(\left(s_{12}+s_{23}\right) z_6-s_{13}+z_8\right).
\end{split}
\ee
The integrand is the square root of a quartic polynomial in $z_8$, with four different roots. This means that the integrand has two genuine branch cuts that cannot be removed by any change of variables, yielding an elliptic function upon integration~\cite{CaronHuot:2012ab,Sogaard:2014jla}.  

For completeness let us also show that localizing the two loops individually gives a consistent result. First we may localize the integration momentum $k_1$ by cutting propagators 1,2,5,6 (using the numbering of (\ref{eq:Family A})).
This yields the result,
\begin{align}
\bar{I}_{\text{box-cut}}& = s_{12}^2 \int \frac{dk_2^4}{(i\pi^{2})^2} \frac{1}{J(k_2)\, \big(m^2-(k_2+p_{12})^2 \big) \, \big(m^2-(k_2-p_3)^2 \big) }\,,
\label{eq:boxcut}
\end{align}
where the Jacobian of the contour deformation $J(k_2)$ reads,
\begin{equation}
J(k_2)= s_{12}^2 \sqrt{s_{12} (s_{12} (2 p_1\cdot k_2+k_2^2+m^2){}^2-4 m^2 (k_2^2 s_{12}-4 p_1\cdot k_2\, p_2\cdot k_2))}\,.
\end{equation}
We note that in the limit $m^2\rightarrow 0 $ the Jacobian reduces to,
\begin{equation}
J(k_2)|_{m^2\rightarrow0} = s_{12}^3 (k_2+p_1)^2\,,
\end{equation}
reproducing the well known result for the cut of the massless case.

Localizing the contour onto the two genuine propagators of  \eqref{eq:boxcut}, will yield an expression similar to  \eqref{eq:hexacut} - an inverse square root of a quartic polynomial with no repeated roots.


\bibliographystyle{JHEP}
\bibliography{refs}






\end{document}